\documentclass[pra,twocolumn,showpacs,preprintnumbers,amsmath,amssymb]{revtex4-1}

\usepackage{graphicx}
\usepackage{dcolumn}
\usepackage{bm}
\usepackage{psfrag}
\usepackage{epsfig}
\usepackage{amsmath}
\usepackage{amssymb}
\usepackage{color}
\usepackage{bbm}
\usepackage[FIGTOPCAP,raggedright,nooneline]{subfigure}
\usepackage{epstopdf}
\usepackage{verbatim}
\usepackage{dsfont}
\usepackage{tabu}

\newcommand{\lk}{\left}
\newcommand{\rk}{\right}

\newcommand{\ignore}[1]{}

\DeclareMathOperator{\arccosh}{arccosh}
\DeclareMathOperator{\cotha}{coth}
\DeclareMathOperator{\tanha}{tanh}
\newcommand{\eq}{Eq.~}
\newcommand{\eqs}{Eqs.~}
\newcommand{\fig}{Fig.~}
\newcommand{\figs}{Figs.~}
\newcommand{\cf} {cf.~}
\newcommand{\ug} {=}

\begin{document}

\title{Atom-field dressed states in slow-light waveguide QED}

\author{Giuseppe Calaj\'o$^1$, Francesco Ciccarello$^2$,  Darrick Chang$^3$ and Peter Rabl$^1$}
\affiliation{$^1$Vienna Center for Quantum Science and Technology,
Atominstitut, TU Wien, Stadionallee 2, 1020 Vienna, Austria}
\affiliation{$^2$NEST, Istituto Nanoscienze-CNR and Dipartimento di Fisica e Chimica, Universita' degli studi di Palermo, Italy}
\affiliation{$^3$ICFO-Institut de Ciencies Fotoniques, The Barcelona Institute of Science and Technology, 08860 Castelldefels (Barcelona), Spain}


\date{\today}

\begin{abstract}
We discuss the properties of atom-photon bound states in waveguide QED systems consisting of single or multiple atoms coupled strongly to a finite-bandwidth photonic channel. Such bound states are formed by an atom and a localized photonic excitation and represent the continuum analog of the familiar dressed states in single-mode cavity QED. Here we present a detailed analysis of the linear and nonlinear spectral features associated with single- and multi-photon dressed states and show how the formation of bound states affects the waveguide-mediated dipole-dipole interactions between separated atoms. Our results provide a both qualitative and quantitative description of the essential strong-coupling processes in waveguide QED systems, which are currently being developed in the optical and the microwave regime.  
\end{abstract}
 

\maketitle

%
%



The coupling of atoms or other emitters to the quantized radiation field can result in drastically different physical phenomena depending on the detailed structure of the electromagnetic  environment. While in free space atom-light interactions are mainly associated with radiative decay, atoms and photons may undergo processes of coherent emission and reabsorption in the case of a single confined mode as studied in the context of cavity QED~\cite{Haroche,Rempe}. Recently, due in part to exciting experimental developments to interface two-level emitters with nanophotonic waveguides~\cite{ReitzPRL2013,YallaPRL2014,ThompsonScience2013,GobanNatComm2014,lodahl, review-lodahl} and to couple superconducting qubits to open transmission lines~\cite{AstafievScience2010,HoiPRL2011,vanLooScience2013,wallraff-superradiance}, 
a different paradigm for light-matter interactions has emerged. \emph{Waveguide QED} refers to a scenario, where single or multiple (artificial) atoms are coupled to a one dimensional (1D) optical channel. 
The 1D confinement of light brings about that individual photons can be efficiently absorbed by even a single atom or  mediate long-range interactions between consecutive atoms along the waveguide. This gives rise to many intriguing phenomena and applications, such as single photon switches and mirrors~\cite{Fan3,lukin-natphys,Nori,Nori2}, correlated photon scattering~\cite{Fan,Baranger1,Gritsev}, self-organized atomic lattices~\cite{ChangPRL2013,GriesserPRL2013}, or the dissipative generation of long-distance entanglement~\cite{GonzalesTudelaPRL2013,moreno,StannigelNJP2014,PichlerPRA2015} and new realizations of quantum gates~\cite{Dzsotjan2010,Baranger2,ciccarello-gate}.

The physics of light-matter interactions in 1D becomes even more involved when the waveguide is engineered to have non-trivial dispersion relations, such as band edges and band gaps~\cite{GobanNatComm2014,review-lodahl,lambro}, near which the group velocity of photons is strongly reduced or free propagation is completely prohibited. In seminal works by Bykov~\cite{Byk}, John and Quang~\cite{John1994} and Kofman {\it et al.}~\cite{Kofman} the decay of an atom coupled to the band edge of a photonic crystal waveguide was shown to exhibit a non-exponential, oscillatory behavior with a finite non-decaying excitation fraction. This behavior can be attributed to the existence of a localized atom-photon bound state with an energy slightly outside the continuum of propagating modes~\cite{John1990,John1991,lambro}.  With many atoms, it has been proposed that the long-lived nature of such states can facilitate the exploration of coherent quantum spin dynamics~\cite{John1996,Douglas2015} or be exploited to engineer long-range photon-photon interactions~\cite{Shahmoon2014,Douglas2015arXiv}.

Motivated by the discussion above, here we study a system of a few quantum emitters, which are coupled to a common `slow-light' photonic waveguide realized by a 1D array of coupled cavities. In the absence of any emitters such a system forms a finite propagating band with an effective speed of light that is fully controlled by the tunnel coupling between neighboring cavities, and thus can in principle be made arbitrarily small.  Coupled cavity arrays (CCA) received large attention, in particular, as a platform for observing quantum phase transitions \cite{CCA_intro,GreentreeNatPhys2006,AngelakisPRA2007,fazio} and for the analysis of photon scattering processes in a finite-bandwidth scenario~\cite{Nori,Sun2,Longo,Longo2,mistery,Schneider,Koc,Roy,Lu,Palma}. Here again the appearance of localized photonic states~\cite{Sun2,mistery,Lu,Palma} results in 
unusual two-photon scattering processes, where, e.g., one photon can remain bound to an atom \cite{Longo,Longo2,mistery}, while the other one escapes. Such processes are absent in free space or infinite-band waveguides.

Building upon those previous findings, we focus in this work specifically on the properties of  dressed atom-photon states, which emerge as the elementary excitations of slow-light waveguide QED systems in the moderate to strong coupling regime. We find that an elegant feature of the CCA system is that in various parameter regimes one can recover the behavior of other systems previously discussed (such as single-mode cavity QED, infinite-bandwidth waveguides and band edges), as well as new phenomena not present in those limiting cases. In our analysis we introduce the single-photon bound states in waveguide QED as continuum generalizations of the dressed-states familiar from the Jaynes-Cummings model for single cavities.  
This analogy then allows us to describe also many properties of the more involved multi-photon and multi-atom settings in terms of the properties of the single-photon dressed state. In particular, we discuss the modefunctions and spectral features of multi-photon dressed states, for which we identify the crossover from a linear regime, where the bound state energies are proportional to the number of excitations, $N_e$, to a nonlinear regime where the splitting of the  bound-state energies from the photonic band scales like $\sim \sqrt{N_e}$. In the last part of the paper, we show how the usual long-range dipole-dipole interactions between multiple atoms coupled to broadband waveguides are modified in the presence of bound photonic states. Here we observe the formation of meta-bandstructures for delocalized dressed states as well as a partial `melting' of these bands back into the continuum, when specific coupling conditions are met. 

The remainder of the paper is structured as follows. In Sec.~\ref{sec:Model} we introduce the basic model of waveguide QED and briefly summarize in Sec.~\ref{sec_weak} the atomic master equation, which describes the dynamics of this system in the weak coupling regime.  In Sec.~\ref{sec:SinglePhoton} we discuss the properties of single photon-bound states in the absence and presence of decay. Finally, in Secs.~\ref{sec:MultiPhoton} and~\ref{multi-atom} we analyze the properties of multi-photon and multi-atom dressed states, respectively. 

\begin{figure}
\includegraphics[width=0.48\textwidth]{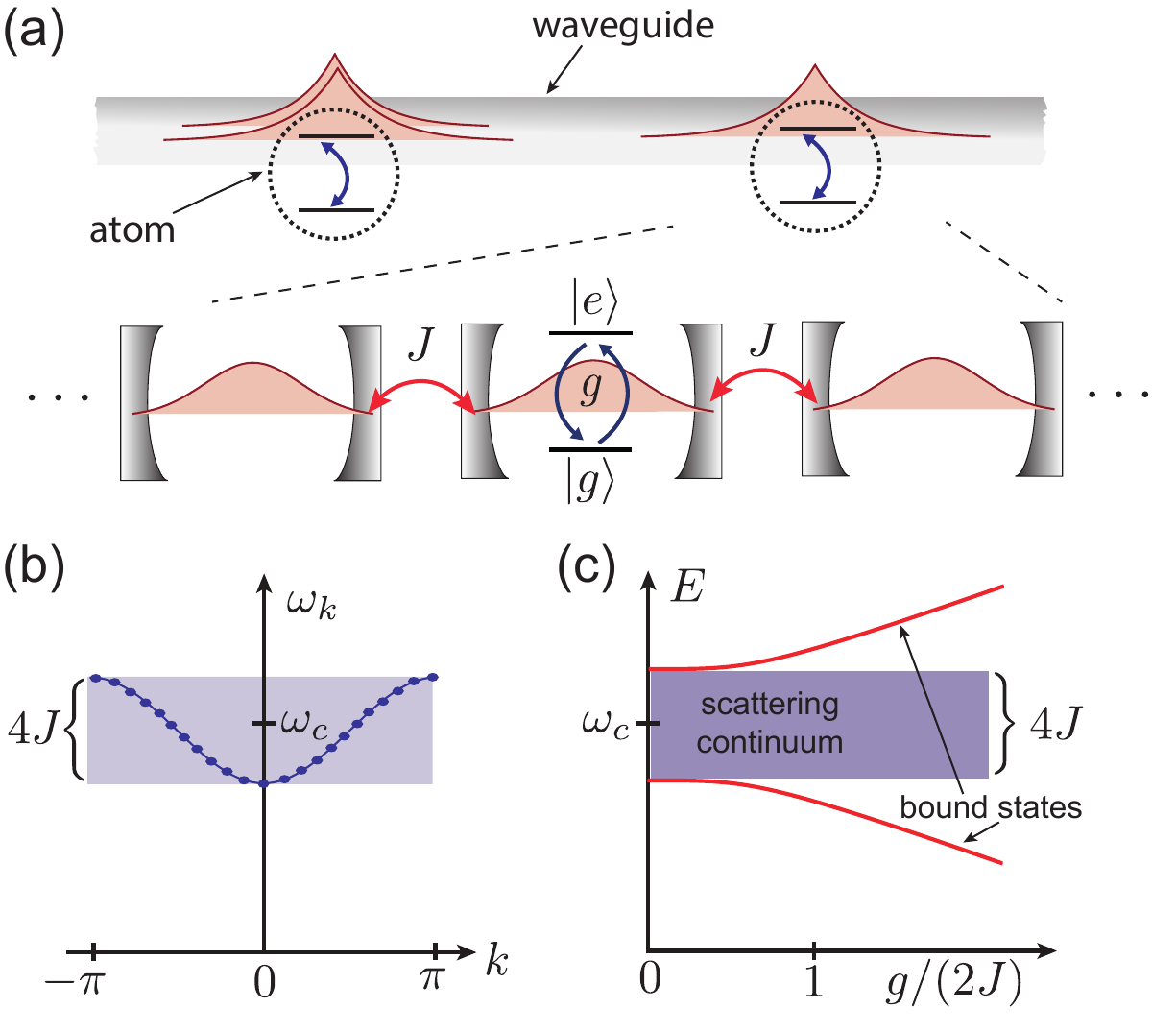}%
\caption{(a) Sketch of a strongly coupled waveguide QED setup with bound atom-photon dressed states around the atomic locations.  The slow-light waveguide can be modelled as a large array of coupled optical resonators with nearest-neighbor coupling $J$. (b) Band structure of the waveguide without atoms. (c) Single-photon (i.e., single-excitation) spectrum as a function of the atom-photon coupling $g$ in the case of a single atom (with $\omega_a=\omega_c$)  coupled to a cavity array according to Hamiltonian \eqref{totalH}. }
\label{Fig1}
\end{figure}

\section{Model}\label{sec:Model}
We consider a system as illustrated in Fig.~\ref{Fig1} (a), where a set of $N_a$ two-level atoms, each with ground (excited) state $|g\rangle$ ($|e\rangle$), are coupled to an optical waveguide of finite bandwidth $4J$. We model the waveguide as an array of $N\rightarrow\infty$ optical resonators with center frequency $\omega_c$ and a nearest-neighbor tunnel coupling $J$. For atoms located at sites $x_i$ the total Hamiltonian for this system is ($\hbar=1$)
\begin{equation}\label{totalH}
\begin{split}
&H= \omega_c \sum_x  a^{\dagger}_x a_x    - J\sum_x(a^{\dagger}_xa_{x-1}+a^{\dagger}_{x-1}a_x)\\
&+ \sum_{i=1}^{N_a} \omega_{a} |e\rangle_i \langle e|  +   g\sum_{i=1}^{N_a} \sum_x\left( a_x \sigma_+^i + a_x^\dag \sigma_-^i\right) \delta_{x,x_i},
\end{split}
\end{equation} 
where $a_x$ ($a_x^\dag$) are bosonic annihilation (creation) operators for the individual cavity modes, $\sigma_-^i = (\sigma_+^i)^\dag =|g\rangle_i\langle e|$, $\omega_a$ is the atomic transition frequency and $g$ is the atom-photon coupling strength. Note that for the validity of Eq.~\eqref{totalH} it has been assumed that $\omega_a\approx \omega_c$ and that both frequencies are much larger than the couplings $g$ and $J$ so that counter-rotating terms can be neglected. Under these assumptions we can eliminate the absolute optical frequencies by changing into a rotating frame with respect to $\omega_c$, and the resulting system dynamics depends only on the atom-photon detuning $\delta=\omega_a-\omega_c$. To account for atomic emission into other radiation modes as well as the absorption of photons in the waveguide, we introduce a bare atomic decay rate $\gamma_a$ and a photon loss rate $\gamma_c$ for each cavity as additional phenomenological parameters. 

The first line of Eq.~\eqref{totalH} represents the tight-binding Hamiltonian $H_c$ of the waveguide. 
By introducing momentum operators $a_k=\frac{1}{\sqrt{N}} \sum_x   e^{i k x}  a_x$, where $k\in \left]-\pi,\pi\right]$, this can be written in a diagonal form $H_c= \sum_k \omega_k a^\dag_k a_k$, with mode frequencies 
\begin{equation}\label{dis}
\omega_k=\omega_c-2J\cos(k),
\end{equation}  
lying inside a band of total width  $4J$ and centered around the bare cavity frequency $\omega_c$ [see Fig.~\ref{Fig1} (b)]. The propagation of photons inside the waveguide is characterized by the group velocity 
\begin{equation}\label{eq:GroupVelocity}
v_{\rm g}(\omega)= \left.\frac{\partial \omega_k}{\partial k}\right|_{\omega_k=\omega}=\sqrt{4J^2-(\omega-\omega_c)^2},
\end{equation}
which vanishes for $J\rightarrow 0$ or when operating at frequencies close to the band edges, i.e., $\omega\approx \omega_c\pm 2J$. In the limit $J=0$ the cavities are completely decoupled, each site being thereby described by a single-mode Jaynes-Cummings model~\cite{Haroche} with coupling constant $g$ and detuning $\delta$. In this sense the present model captures well finite-bandwidth and bandedge features  over a
 wide range of parameters. However, note that the CCA  may only crudely approximates the actual dispersion relation in real photonic bandstructures and does not include effects like directional emission, which can occur in certain waveguide implementations~\cite{PetersenScience2014,MitschNatComm2014,SollnerNatNano2015}.

\section{Broadband limit}\label{sec_weak}

Let us first consider the weak-coupling or broadband limit  $g/J\rightarrow 0$. In this regime the photonic waveguide modes simply act as a collective reservoir for the atoms and can be eliminated by using a Born-Markov approximation.  As a result we obtain a master equation for the reduced density operator of the atoms (see App.~\ref{app:ME}) 
\begin{equation}\label{mastergen}
\dot \rho = -i [H_a, \rho] + \sum_{i,j}\frac{\Gamma_{ij}}{2} \left(2 \sigma_-^j\rho \sigma_+^i -  \sigma_+^i\sigma_-^j\rho -  \rho \sigma_+^i\sigma_-^j\right),
\end{equation} 
where in the rotating frame with respect to $\omega_c$, 
\begin{equation}\label{Hmaster}
H_a= \sum_{i} \delta  |e\rangle_i \langle e|    + \frac{1}{2} \sum_{i,j}  U_{ij} \left( \sigma_+^i\sigma_-^j + \sigma_-^i\sigma_+^j\right). 
\end{equation}
In \eqs (\ref{mastergen}) and (\ref{Hmaster}) the $\Gamma_{ij}$ and $U_{ij}$ represent correlated decay rates and coherent dipole-dipole interactions, respectively,  which arise from virtual or real photons propagating along the waveguide.
\begin{figure}
\includegraphics[width=0.48\textwidth]{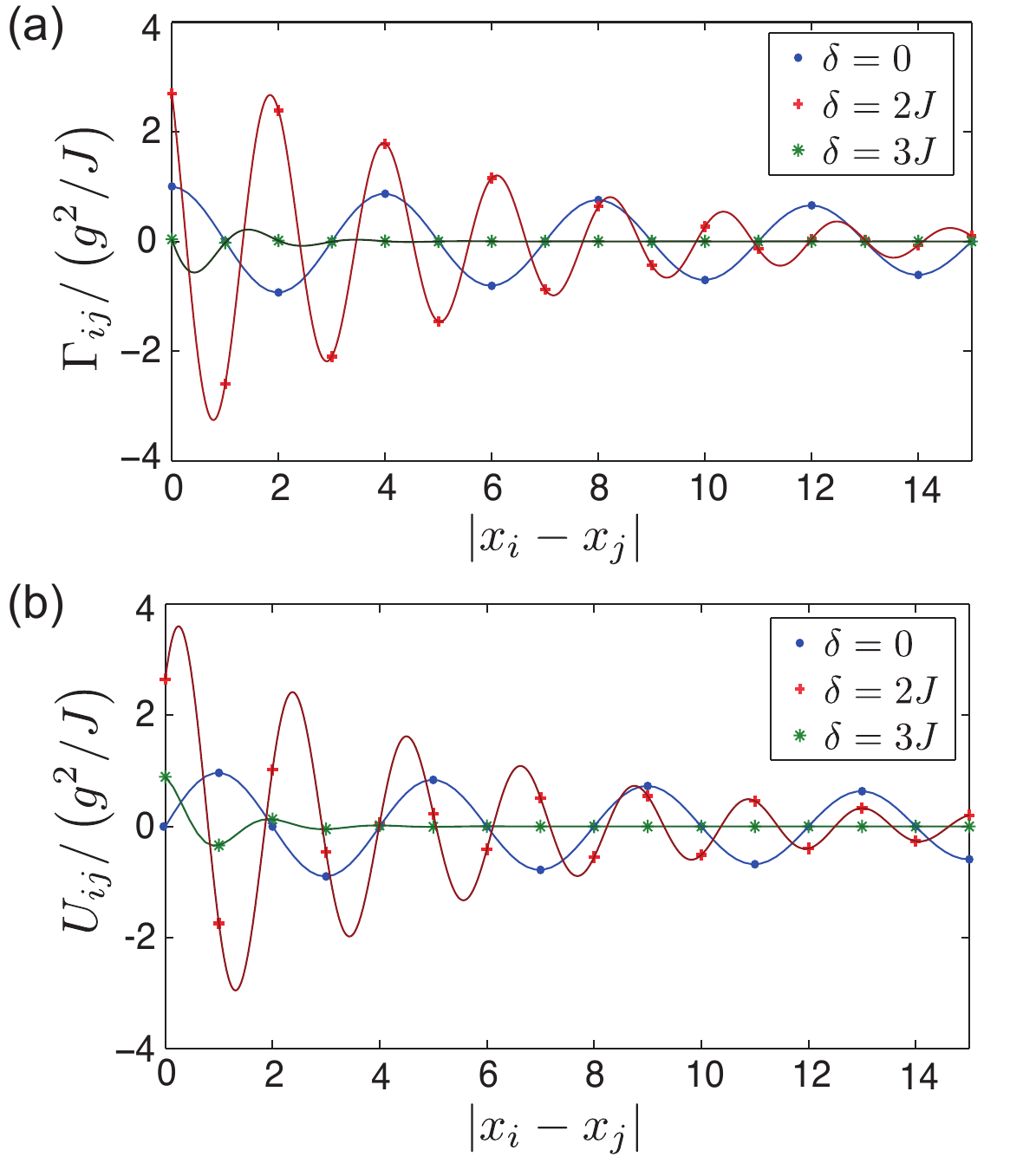}%
  \caption{(a) Correlated decay rates $\Gamma_{ij}$ against the (discrete) interatomic distance $|x_i-x_j|$ and (b) coherent dipole-dipole interactions $U_{ij}$ versus $|x_i-x_j|$ for different detunings $\delta=\omega_a-\omega_c$.
The solid lines are a guide to the eye obtained from a continuous interpolation of Eq.~\eqref{eq:Aij}. In each case, the photon loss rate has been set to $\gamma_c/(2J)=0.14$.}
  \label{rates}
\end{figure}
By taking into account small atomic and photonic losses
we obtain $\Gamma_{ij}=2{\rm Re}\{ A_{ij}\}+\gamma_a$ and $U_{ij}=2{\rm Im}\{ A_{ij}\}$,
where
\begin{equation}\label{eq:Aij}
\begin{split}
A_{ij}&= \frac{g^2 }{\tilde v_{\rm g}(\delta)}  e^{ i K   |x_i-x_j]},\\
\end{split}
\end{equation}
and
\begin{equation}\label{eq:K}
K= \pi -\arccos\left[{\frac{\delta+i\gamma_c/2}{2J}}\right].
\end{equation}
Here we have introduced a generalized (complex) group velocity
\begin{equation}\label{eq:tildevg}
\tilde v_{\rm g}(\delta)=\sqrt{4J^2-\left(\delta+i\frac{\gamma_c}{2}\right)^2}.
\end{equation}
For $\gamma_c\rightarrow 0$ and for atomic frequencies within the photonic band this quantity reduces to the conventional group velocity given in Eq.~\eqref{eq:GroupVelocity}. In this case $\sim 1/|v_g(\delta)|$ determines the density of photonic modes, or equivalently, the correlation time of the waveguide.  
In a system with losses this correlation time is now replaced by $1/|\tilde v_{\rm g}(\delta)|$, which is well-defined and non-diverging even at or beyond the band edges (for a related study of the group velocity in lossy waveguides see also Ref.~\cite{PedersenPRB2008}). Therefore, the Born-Markov approximation, which requires
\begin{equation}\label{eq:BornMarkov}
g\ll  |\tilde v_{\rm g}(\delta)|,
\end{equation}
can be used for all atomic frequencies provided that the coupling $g$ is sufficiently weak and photon propagation times are negligible (see App.~\ref{app:ME} for additional details on the validity of the Born-Markov approximation).

Figure~\ref{rates} illustrates the dependence of $\Gamma_{ij}$ and $U_{ij}$ on the interatomic distance for different atom-photon detunings, $\delta$, and a non-vanishing photon loss rate $\gamma_c$. If instead cavity losses are negligible, Eqs.~\eqref{Hmaster}-\eqref{eq:K} reproduce the effective spin model for two-level atoms coupled to an infinite-bandwidth waveguide~\cite{ChangNJP2012,GonzalesTudelaPRL2013}. In particular for frequencies within the propagating band, $K$ becomes purely real and the system thus supports both coherent and dissipative dipole-dipole interactions of equal strength,
\begin{equation}\label{Gij}
\begin{split}
\Gamma_{ij}& \simeq \frac{2g^2}{v_g(\delta)} \cos\left( K|x_i-x_j|\right), \\
 U_{ij} &\simeq \frac{2g^2}{v_g(\delta)} \sin\left( K|x_i-x_j|\right).
\end{split}
\end{equation}
This coupling is infinite in range, with a phase factor $e^{iK|x_i-x_j|}$ that reflects the propagation phase of photons at the atomic resonance frequency that mediate the interaction. This behavior can be seen in Fig. \ref{rates} for $\delta = 0$ (blue curve), with the deviation from infinite-range interaction due to the finite cavity losses $\gamma_c$.  As expected, by going from the center of the band towards the edge, $\delta \approx 2J$ (red curve), both the coherent couplings as well as the correlated decay rates increase due to a reduction of the group velocity. However, slow propagation also means that the photons have more time to decay and for a finite $\gamma_c$ and large atom-atom distances, there is a trade-off between an enhanced coupling and a larger propagation loss. For atomic frequencies outside the band there are no longer waveguide modes into which the atom can emit. Therefore, for $\gamma_c\rightarrow 0$, the real part of $A_{ij}$ vanishes and the atoms interact predominantly in a coherent way via a virtual exchange of photons. The exponential decay of interactions directly reflects the  exponential attenuation of fields propagating through a band gap (see green curve of Fig.  \ref{rates}).

In summary Eq.~\eqref{mastergen} shows that for sufficiently weak coupling the dynamics of the waveguide QED system can be described in terms of atomic excitations, which interact via a quasi-instantaneous exchange of photons. In this regime it is preferential to work near the band edge or to reduce the waveguide bandwidth all together in order to enhance waveguide mediated atom-atom interactions (coherent or dissipative) compared to the bare atomic decay. However, eventually the Markov condition given by Eq.~\eqref{eq:BornMarkov} breaks down and for larger couplings the photons emitted by an atom can be coherently reabsorbed before they decay or propagate along the fiber. In this strong coupling regime photons and atoms can be bound together and form new hybridized excitations.

\section{Atom-photon dressed states}\label{sec:SinglePhoton}  
In the absence of other decay channels, the atom-light coupling in Eq.~\eqref{totalH} conserves the total number of photons and excited atoms, $N_e=\sum_x a_x^\dag a_x {+} \sum_i |e\rangle_i\langle e_i|$,  and the eigenstates of $H$ can be discussed separately within each subspace of given excitation number. For a given value $N_e$, the Schr\"odinger equation $H|\phi\rangle=E|\phi\rangle$ then has two types of solutions. First, there are scattering states, which are spatially extended over the whole waveguide and have an energy $E/N_e\in [-2J,2J]$ within the free $N_e$-photon band.  Second, there are states with energy  $|E|/N_e > 2J$~\cite{Footnote1}. These states are energetically separated from the $N_e$-photon continuum and represent bound states with an exponentially localized photonic component. While both types of states are atom-photon dressed states, in this work we are primarily interested in the latter type, namely in {\it bound} dressed states.   
 
Note that in the waveguide QED literature the term
{\it photon bound state} is also used to describe correlated propagating
multi-photon wavefunctions scattered by a nonlinear emitter.
These states are not localized around the atom but they are typically infinite in spatial extent, and bound only with respect to the relative coordinates of the photons~\cite{Fan}. In this paper we do not consider this kind of states and use the term {\it bound state} only for wavefunctions
spatially localized around the atomic position.
\begin{figure}
\includegraphics[width=0.48\textwidth]{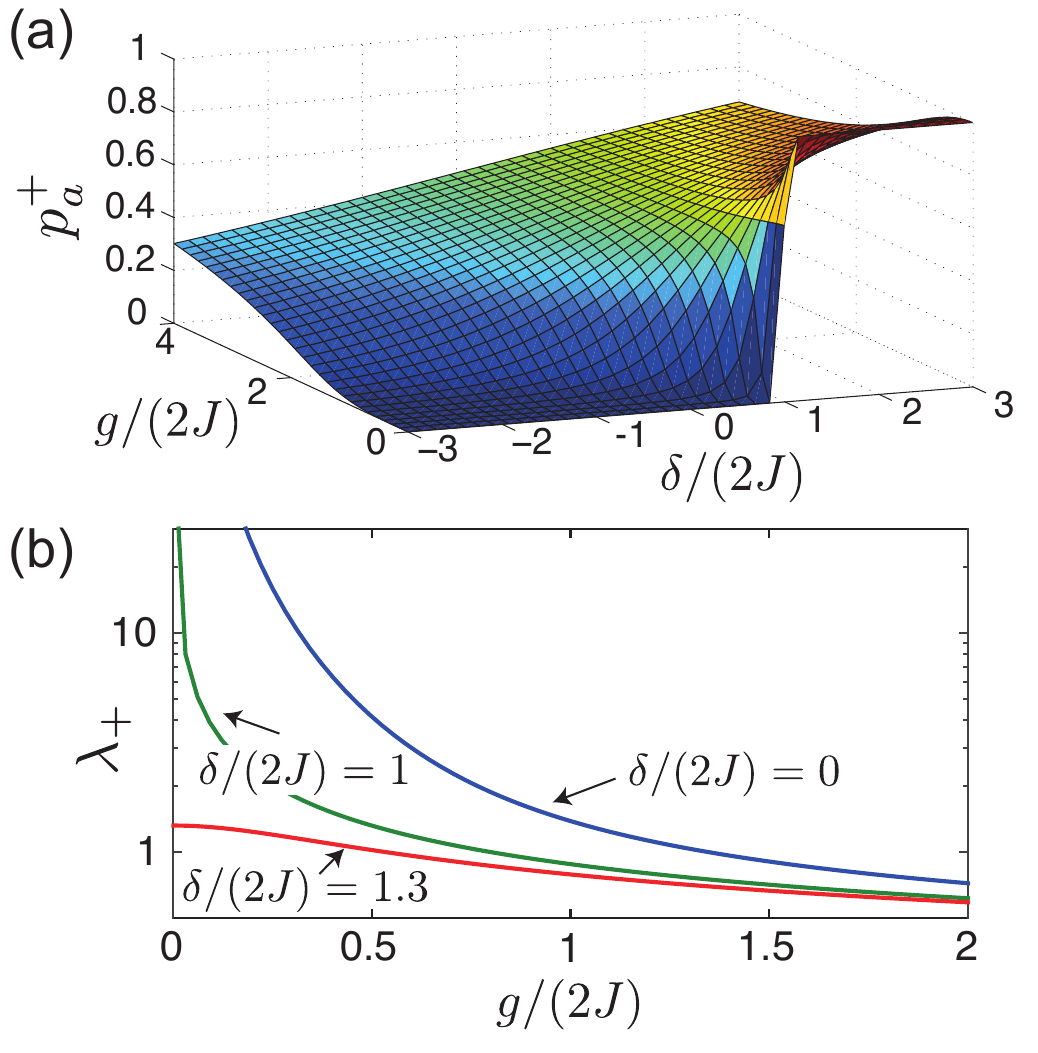}%
  \caption{(a) Atomic  population $p_a^+=\cos^2(\theta_+)$ in the upper bound state as function of the coupling constant $g$ and the atom-field detuning $\delta$. (b) The width of the photonic wavepacket in the upper bound state, $\lambda_+$, is plotted as a function of $g$ and for three different detunings $\delta$.}
\label{figpop}
\end{figure}

 \subsection{Single-photon dressed states}\label{sec1at}
 We first consider the simplest setting of a single photon coupled to a single atom located at position $x_a$. In this case, $N_e{=}1$ and the solutions of the Schr\"odinger equation $H|\phi\rangle=E|\phi\rangle$ are superpositions of an atomic excitation $|e,0\rangle$ and single photon states $|g,1_x\rangle\equiv a_x^\dag |g,0\rangle$ ($|0\rangle$ is the field vacuum state). Figure \ref{Fig1} (c) shows the resulting energy spectrum which consists of the above mentioned band of scattering states and two bound states with energies $E_{\pm}$, which are the real solutions of (see App.~\ref{appB})
\begin{equation}\label{eq:Ebound}
E_{\pm}-\delta=\frac{g^2}{E_{\pm}\sqrt{1-\frac{4J^2}{E^2_{\pm}}}}\,\,.
\end{equation}
The corresponding bound-state wavefunctions can be written in the form
 \begin{equation}\label{boundcav}
 \begin{split}
|\phi_{\pm}\rangle=& 
\left[ \cos{\theta_{\pm}} \sigma_+ \pm  \sin{\theta_{\pm}}   a_{\lambda,\pm}^\dag(x_a)  \right] |g,0\rangle \equiv D^\dag_{\pm} (x_a)|g,0\rangle.
\end{split}
\end{equation} 
Here we have defined the normalized bosonic creation operator 
\begin{equation}\label{alop}
a_{\lambda,\pm}^\dag(x_a) =   \sum_x\frac{ (\mp1)^{|x-x_a|} e^{-\frac{|x-x_a|}{\lambda_\pm}} }{{\sqrt{\coth{\frac{1}{\lambda_\pm}}}}} a^\dag_x,
\end{equation}
which creates a photon in an exponentially localized wavepacket around the atom's position $x_a$. In Eqs. (\ref{boundcav}) and (\ref{alop}) the size of the photonic wavepacket, $\lambda_\pm=\lambda(E_{\pm})$, and the mixing angle $\theta_\pm=\theta(E_{\pm})$ are functions of the corresponding bound state energies.  These two parameters determine the nature of the bound state wavefunctions and are given by
\begin{equation}\label{tetha}
\cos  \theta =\left(1+\frac{g^2}{E^2\left(1-\frac{4J^2}{E^2}\right)^{\frac{3}{2}}}\right)^{-\frac{1}{2}}\mbox{,}
\end{equation} 
and
\begin{equation}\label{lambda}
\frac{1}{\lambda}=\arccosh{\left(\frac{|E|}{2J}\right)}.
\end{equation} 
Figure~\ref{figpop} summarizes the dependence of $\lambda_+$ and the atomic excited-state population $p_a^+=\cos^2(\theta_+)$ on the coupling $g$ and the atom-photon detuning $\delta$. The analogous quantities associated with $E_-$ can be inferred through the identities $\lambda_-(\delta)=\lambda_+(-\delta)$ and $\theta_-(\delta)=\theta_+(-\delta)$.

\emph{Discussion.} In their respective limits, Eqs.~\eqref{eq:Ebound}-\eqref{lambda} reproduce various results that have been previously obtained for photonic bound states near band edges or in coupled cavity arrays \cite{John1990,John1991,lambro,Sun2,Longo2,mistery,Lu,Palma}.  The form of the wavefunction given in Eq.~\eqref{boundcav} provides a unified description of all those cases in terms of the mixing angles $\theta_\pm$ and the wavepacket lengths $\lambda_\pm$. It also establishes a direct connection to the more familiar dressed states of the single mode Jaynes-Cummings model~\cite{Haroche} by taking the limit $J\rightarrow 0$, where 
$E_{\pm}=\frac{\delta}{2}\pm\frac{1}{2}\sqrt{\delta^2+4g^2}$, $\theta_+=\theta_--\pi/2$ and $\lambda_\pm\approx 0$. For a finite $J$  this single-cavity  picture is modified in two ways. First, the photonic component now extends over multiple sites and becomes more and more delocalized the weaker the coupling $g$. Second, the total atomic contribution to both bound states, $\cos^2(\theta_+)+\cos^2(\theta_-) < 1$, is always smaller than one and for $|\delta|<2J$ it vanishes as $g/J\rightarrow 0$. Although a bound state solution always exists, both dressed states become more photon-like as $g/J$ decreases and eventually become indistinguishable from the propagating waveguide modes. For atomic frequencies outside the band, e.g., $\delta>2J$, the upper bound state becomes more atom-like as $g/J\rightarrow 0$, but the residual photonic cloud remains localized. Overall, these results show that a simplified model where the waveguide is replaced by an effective cavity of size $\lambda$ would be incomplete.  In particular, such a description misses the fact that for $\delta\neq 0$ photonic wavefunctions associated with the two dressed states can significantly differ, i.e., $\lambda_+\neq\lambda_-$ and $\theta_+\neq\theta_-$.

\subsection{Excitation spectrum}
An experimentally relevant quantity to probe the properties of atom-photon dressed states is the atomic excitation spectrum $S_a(\omega)$, which can be obtained by weakly exciting the atom with a laser of frequency $\omega$ and recording the total emitted light. In the weak driving limit   
the excitation spectrum is given by 
\begin{equation}
S_a(\omega)= \frac{\gamma_a^2}{4} \left| \langle e,0|  \frac{ 1}{H_{\rm eff} - \omega \mathbbm{1} }  |e,0 \rangle\right|^2,
\end{equation}
where
\begin{equation}\label{Heff}
H_{\rm eff} = H -i \frac{\gamma_a}{2} |e\rangle\langle e| - i  \sum_x \frac{\gamma_c}{2} a^\dag_x a_x,
\end{equation}
and the normalization has been set such that $S(\omega=\omega_a)=1$ for $g=0$. 
\begin{figure}
\includegraphics[width=0.5\textwidth]{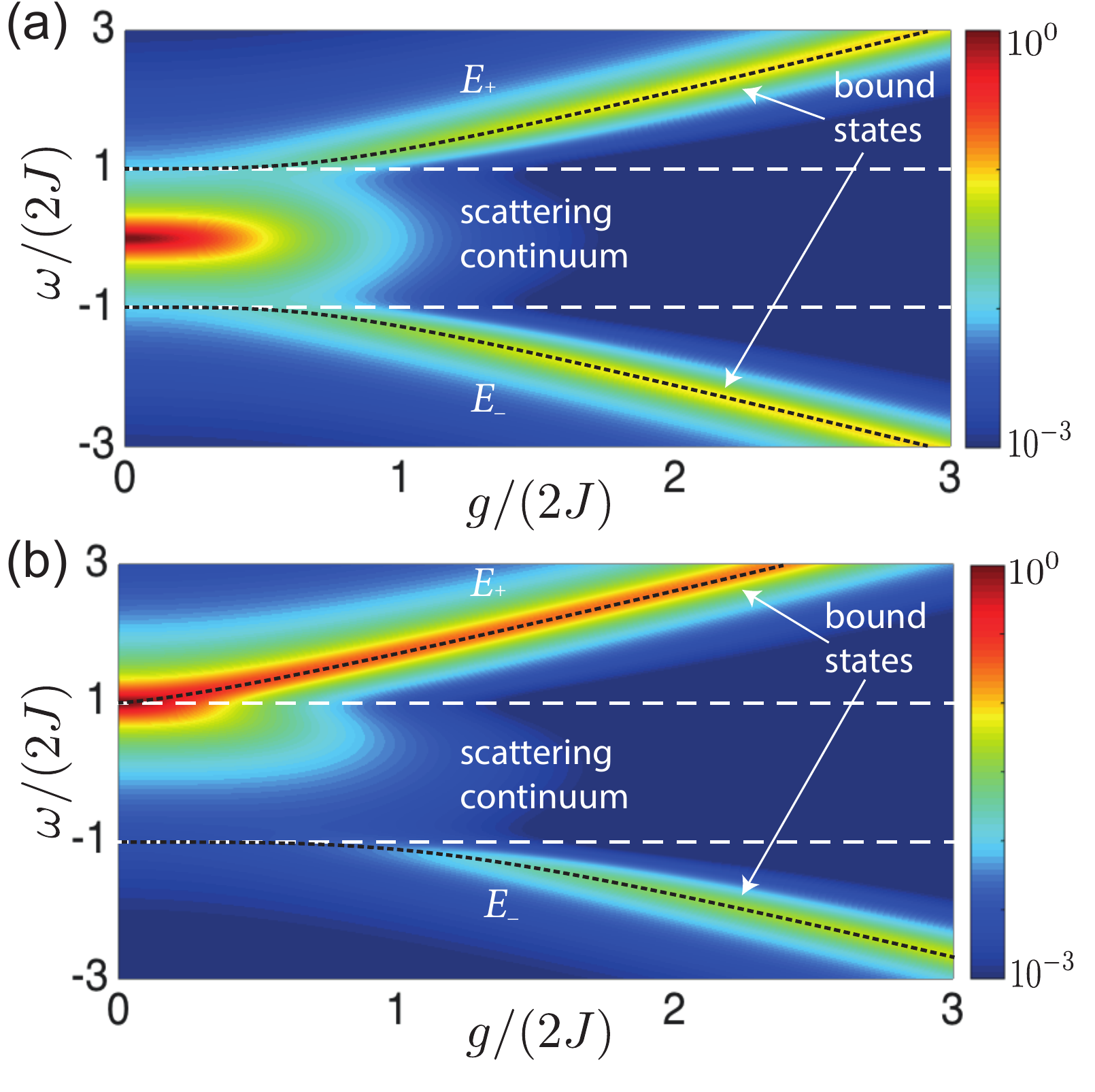}%
\caption{Atomic excitation spectrum $S_a(\omega)$ (in logarithmic scale) as function of $g$ and for an atom-cavity detuning (a) $\delta=0$ and (b) $\delta=2J$. The dotted lines show the bound-state energies $E_{\pm}$ in the absence of loss, while the dashed lines correspond to the waveguide band edges. In either case, we have set $\gamma_a/(2J)=0.1$ and $\gamma_c/(2J)=0.2$.}
  \label{decayspectra}
\end{figure}
 Figure~\ref{decayspectra} shows the results for  $S_a(\omega)$  for different coupling strengths $g$ and for the two relevant cases $\delta=0$ (center of the band)  and $\delta=2J$ (upper-band edge). For $\delta=0$ we observe three different regimes. For very weak coupling there is only a single peak at the atomic frequency with a width $\sim \gamma_a+ g^2/J$ due to the enhanced emission into the waveguide (recall that in the broadband limit the atom emission rate into the waveguide is $g^2/J$, see Section \ref{sec_weak}).  At intermediate couplings $g/(2J)\sim1$  the spectrum is completely smeared out. The atom is now partially hybridized with all waveguide modes and there is no longer a well defined frequency associated with the atomic excitation. At larger couplings two dominant resonances at the dressed-state energies $E_{\pm}$ appear. As the coupling increases the width of the two bound-state resonances approaches 
 \begin{equation}\label{gamma-av}
\bar \gamma= \frac{\gamma_a+\gamma_c}{2},
\end{equation}
as expected from an equal superposition of atomic and photonic excitations. 
For $\delta = 2J$ a significant hybridization between atom and photon is already observed at small $g$, consistent with the atomic population $p_a^+ \approx 0.67$ predicted for the dressed state exactly at the band edge [see Fig.~\ref{figpop} (a)] . However, in this case the transition from waveguide-enhanced decay to atom-photon hybridization is not apparent and will be discussed in more detail in the following.

\subsection{Onset of strong coupling}
An important regime of operation in cavity QED is the regime of strong coupling, where the coherent interaction  between atoms and photons dominates over the relevant decay processes. For a single cavity in resonance with the atom this regime is usually defined by the condition 
\begin{equation}
g >  \frac{\gamma_a+\gamma_c}{4}.  
\end{equation}
Our goal is now to identify an equivalent condition for the waveguide QED system, by taking a closer look at the spectral features for $g\ll J$. Note that  the atomic excitation spectrum is in general given by \cite{John1994,Schneider}
\begin{equation}\label{specself}
S_a(\omega)= \frac{\gamma_a^2}{4} \frac{  1}{  \left| \omega-\delta+i\frac{\gamma_a}{2}-\tilde  \Sigma(\omega)\right|^2},
\end{equation}
where $\tilde \Sigma(\omega)=-i g^2/ \tilde v_{\rm g}(\omega)$ is the self energy in the presence of dissipation. To bring this result into a more useful form we define 
\begin{equation}
\Delta_{\pm}(\omega)=\left( \omega-\delta+i\frac{\gamma_a}{2}\right)\tilde v_{\rm g}(\omega)\pm ig^2.
\end{equation}
%
It can be shown that $\Delta_+(\omega)\Delta_-(\omega)$ is a forth order polynomial in $\omega$ with two roots given by the complex eigenenergies $\tilde E_{\pm}$ of $H_{\rm eff}$.
%
%
We can use this property to further rewrite the spectrum as 
\begin{equation}\label{spec3}
S_a(\omega)= \frac{\gamma_a^2}{4}  \frac{ | \tilde v_{\rm g}(\omega)\Delta_{-}(\omega)|^2}{|(\omega-\tilde E_+)(\omega-\tilde E_-)L(\omega)|^2}.
\end{equation}
Here $L(\omega)$ is a quadratic polynomial, which for the limits discussed below has two roots with real parts inside the photonic band, and thus describes the atomic emission into the waveguide continuum.  Overall the structure of the spectrum then consists of two external poles with a position and a width given by the real and imaginary parts of $\tilde E_\pm$ and a broader emission peak inside the waveguide. Note that for $\gamma_c\rightarrow 0$ the generalized group velocity, $\tilde v_g(\omega)$, and therefore also the spectrum vanishes exactly at the bandedge, $\omega=\pm 2J$.  This is due to a destructive interference between the excitation laser and the long-lived band-edge mode and leads to a Fano-like profile for $S_a(\omega)$. For non-vanishing $\gamma_c$ this interference effect is washed out. 

\begin{figure}
\includegraphics[width=0.49\textwidth]{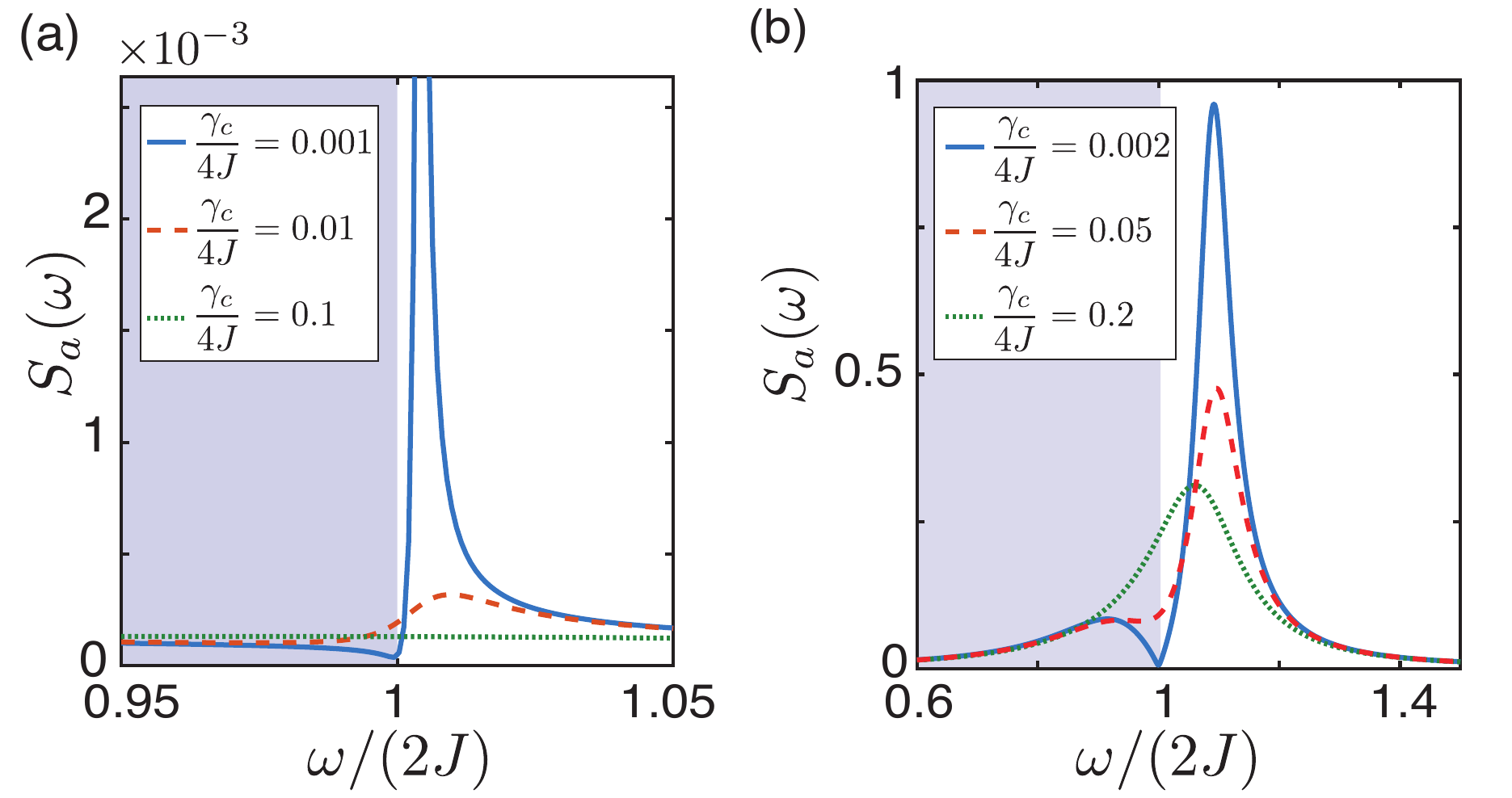}%
\caption{Dependence of the atomic excitation spectrum $S_a(\omega)$ near the band edge and for  (a) $\delta=0$ and (b) $\delta=2J$. In (a) the values $g/(2J)=0.3$ and $\gamma_a/(4J)=0.02$ and in (b) the values $g/(2J)=0.2$ and $\gamma_a/(4J)=0.05$ have been assumed and in both cases the spectrum is plotted for different cavity decay rates $\gamma_c$. 
%
}
  \label{sp_edge}
\end{figure}

We first consider the case $\delta=0$, where  we obtain to lowest order in $g$ 
\begin{equation}\label{apprpole}
\begin{split}
& \tilde E_{\pm}\simeq \pm 2J\pm\frac{g^4}{16J^3[1\pm i(\gamma_a-\gamma_c)/(2J)]} -i \frac{\gamma_c}{2},
 \end{split}
\end{equation}
which shows that for not too large decay rates, the position of the external peaks essentially follows the bare energy levels $E_\pm$ and their width is mainly determined by photon loss.  For the polynomial determining the internal peaks we obtain
\begin{equation}
L(\omega)=\left(\omega+i\frac{\gamma_a}{2}\right)^2+\left(\frac{g^2}{2J}\right)^2,
\end{equation}
which therefore contributes with two purely imaginary poles at $\omega=- i(\gamma_a \pm g^2/J)/2$. Figure~\ref{sp_edge} (a)  shows a zoom-in on the resulting spectrum near the band edge and for different values of $\gamma_c$. First, we observe that for large $\gamma_c$ the external peak is completely buried within the tail of the broad internal peak and a closer inspection shows that a minimal coupling of 
 \begin{equation} \label{eq:SCC1}
 g> \sqrt{J\gamma_c},
 \end{equation}
is required to spectrally resolve the existence of an external bound state. This condition is equivalent to the requirement that the atomic emission rate into the waveguide exceeds the cavity loss rate. Once this condition is fulfilled we can define strong coupling by the requirement that the separation of the external peak from the band edge, ${\rm Re} \{\tilde E_+-2J\}$, exceeds its half-width given by ${\rm Im} \{\tilde E_+\}$. Again for $\gamma_c,\gamma_a\ll 2J$ we obtain 
\begin{equation}\label{eq:SCC2}
g>\sqrt[4]{8J^3\gamma_c},
\end{equation}
as the strong coupling condition for a resonantly coupled waveguide QED system. Note that since in the present regime  the bound states are mainly photonic in nature the atomic decay is relevant only for higher-order corrections.

The second important limit is $\delta=2J$, which for $g/J\ll 1$ also corresponds to the  quadratic dispersion relation assumed in studies of photonic bound states near the band edge of a photonic crystal waveguide \cite{John1990,John1991,John1994,Kofman,lambro}. Note that in this regime the initial scaling of  the bound state energy in the absence of losses ($\gamma_a=\gamma_c=0$) is given by
\begin{equation}\label{2JEc}
E_{+}\simeq 2J+\left(\frac{g^4}{4J}\right)^{\frac{1}{3}}\mbox{,}
\end{equation}
where the splitting $\beta=\sqrt[3]{g^4/(4J)}$ can be directly identified with the frequency of coherent atom-photon oscillations at the band edge~\cite{John1991}.
In the presence of decay and for $g< |\gamma_c-\gamma_a|$ we obtain instead the modified result  
\begin{equation}\label{2Jkg}
\tilde E_{+}\simeq2J-i\frac{\gamma_a}{2} +\frac{g^2}{2\sqrt{J|\gamma_c-\gamma_a|}}(1\mp i)\mbox{,}
\end{equation}
where the minus (plus) sign is for the case $\gamma_c>\gamma_a$ ($\gamma_c<\gamma_a$). This result shows that not only does the presence of loss modify the initial scaling of the bound state energy,  Eq.~\eqref{2Jkg} also predicts that at the band edge and  for small $g$ the atom is \emph{critically} damped, i.e., the coupling induced losses are exactly of the same magnitude as the coherent shift of the bound state energy. By increasing the coupling further the imaginary part of the eigenvalue $\tilde E_{+}$ will eventually saturate at a value $\bar \gamma /2$ [\cf\eq(\ref{gamma-av})] corresponding to a fully hybridized state. This hybridized regime is reached for coupling strengths
\begin{equation}\label{eq:SCC3}
g > \sqrt[4]{\frac{J|\gamma_c-\gamma_a|^3}{4}}\,.  
\end{equation}
Under this condition the separation of the bound-state from the bandedge is then given by $\beta$ from which we obtain the strong coupling condition $\beta>\bar \gamma/2$, or 
\begin{equation}\label{eq:SCC4}
g > \sqrt[4]{ J \bar \gamma^3/2}\,.  
\end{equation}
Figure \ref{sp_edge} (b) shows a zoom-in of the atomic spectrum $S_a(\omega)$  for $\delta=2J$ and for three different values of the photon decay, which correspond to the critically damped, intermediate and strong coupling regime. Note that for $\delta=2J$ the internal poles associated with $L(\omega)$, i.e., $\omega_1= 2J-i\frac{\gamma_c}{2}$  and $\omega_2= 2J-i\frac{\gamma_a}{2}-g^2(1\mp i)/(2\sqrt{J|\gamma_c-\gamma_a|})$ provide an additional background, but do not play a significant role.

\subsection{Localization} 

For the remainder of this work we are mainly interested in coherent effects and for the sake of clarity we will only present results for idealized systems where $\gamma_a=\gamma_c=0$. Therefore, the validity of these results in particular requires that the strong-coupling conditions identified in Eqs.~\eqref{eq:SCC1}, \eqref{eq:SCC2}, \eqref{eq:SCC3} and \eqref{eq:SCC4}  are fulfilled in the respective limits.  
In addition, it is important to emphasize that all the results discussed in this work are based on the model of a perfectly regular cavity array. In real systems disorder in the cavity frequencies or tunnel couplings introduces an additional localization mechanism, even in the absence of the emitters. To estimate this effect we can consider a simple impurity model, where we add an energy offset $\epsilon$ to one of the lattice sites,  $H_c\rightarrow H_c+\epsilon a^\dag_{x_d} a_{x_d}$. This model is well known in literature \cite{Eco} and it exhibits a purely photonic bound state with a localization length  
\begin{equation}
\frac{1}{\lambda_\ell}= {\rm arcsinh}\left(\frac{|\epsilon|}{2J}\right).
\end{equation}
This means that random energy offsets of typical strength $\epsilon$ will create bound states that are localized over $ \lambda_{\ell}\sim 2J/|\epsilon|$ lattices sites. While atom-photon bound states will also exist in such disordered waveguides, all the  predictions 
in this work are based on the assumption that $\lambda_\ell$ is large compared to the size of the atom-induced bound states, $\lambda_\pm$. For a more accurate treatment of localization in waveguides, see, for example Ref.~\cite{Hern} and the supplementary material of \cite{Douglas2015}.


\section{Multi-photon dressed states}\label{sec:MultiPhoton}
While in cavity QED the appearance of a normal-mode splitting (corresponding to the well-known vacuum Rabi frequency) signifies the onset of strong light-matter interactions, the hallmark of a fully quantized radiation coupling lies in the non-linear scaling of this splitting with the number of excitations, $\sim g\sqrt{N_e}$. In this section we will address the properties of multi-photon dressed states to see to what extent this quantum signature prevails in the context of waveguide QED. In contrast to the single excitation case, the Schr\"odinger equation $H|\psi\rangle=E|\psi\rangle$ for $N_e >1$  no longer permits simple analytic solutions and for exact results one is restricted to numerical methods in real or momentum space \cite{Schneider,Longo,Longo2,mistery}. In this work we perform such calculations by an approximate variational approach, which provides additional intuition on the nature of the multi-photon dressed states, and allows us to evaluate the corresponding bound-state energies for excitation numbers that are no longer trackable by standard numerical methods.

\subsection{Two-photon dressed states}\label{2secp}
 \begin{figure}
 \includegraphics[width=0.48\textwidth]{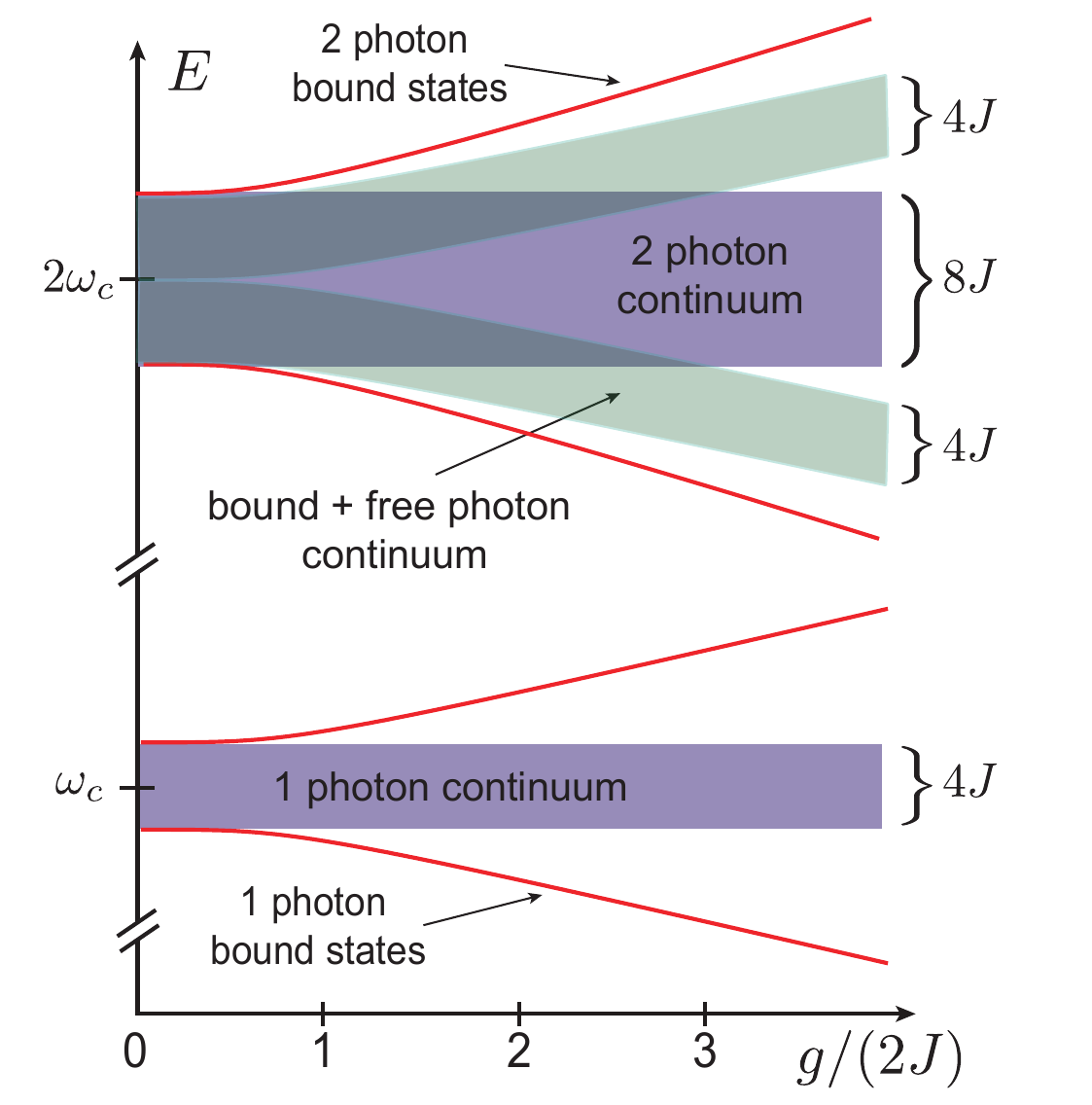}
 \caption{Sketch of the single- and two-excitation spectrum in a finite-bandwidth waveguide coupled to an atom for $\delta=0$. See main text for more details. }\label{2photspfig}
 \end{figure}
Let us first consider the two-excitation subspace, where a general eigenfunction of Hamiltonian (\ref{totalH}) can be written in the form
\begin{equation}\label{2pho}
|\phi\rangle=\sum_xb(x)a_x^{\dagger}|e,0\rangle+\frac{1}{\sqrt{2}}\sum_{x,y} u(x,y)a_{x}^{\dagger}a_{y}^{\dagger}|g,0\rangle\mbox{.}
\end{equation}
By assuming that the atom is located at $x_a=0$ the inversion symmetry of the Hamiltonian and the bosonic symmetry of the wavefunction require $u(x,y)=u(y,x)$, $u(-x,y)=u(x,y)$ and $b(-x)=b(x)$.
This ansatz leads to the set of coupled equations 
\begin{equation}\label{eqond}
\begin{split}
&-J\left[u(x+1,y)+u(x-1,y)+u(x,y+1)\right.\\
&\left.+u(x,y-1) \right]
+\frac{g}{\sqrt{2}}\left[b(x)\delta_{0,y}+b(y)\delta_{0,x}\right]=Eu(x,y)\mbox{,}
\end{split}
\end{equation}
and
\begin{equation}\label{eqond2}
\begin{split}
&-J\left[b(x+1)+b(x-1)\right]\\
&+\frac{g}{\sqrt{2}}\left[u(0,x)+u(x,0)\right]=Eb(x)\,\mbox{.}
\end{split}
\end{equation}
These equations, which extend the continuous waveguide \cite{Fan} case to a discrete model, can be solved numerically and the resulting eigenvalues spectrum is shown in Fig.~\ref{2photspfig} together with the single excitation energy band discussed in Sec.~\ref{sec1at}. For our numerical calculations an array of $N=120$ coupled resonators with periodic boundary conditions has been assumed. 
In line with the single-excitation case, we observe a band of two-photon scattering states with energies $E\in [-4J,4J]$. In addition, there are two bands with energies $E\in [E_{\pm}-2J,E_\pm +2J]$. These bands  can be simply interpreted as the combination of a single-atom bound state with energy $E_\pm$ and an additional free photon with energy $\omega_k$. Finally, we observe two individual lines at energies $E_{\pm}^{(N_e=2)}$ above and below all other states, which represent the true two-photon bound states in the $N_e=2$ sector. 

Before proceeding with a more detailed discussion on the two-photon bound states, let us briefly point out another interesting feature in Fig.~\ref{2photspfig} in the two-excitation manifold, namely the overlap region between the continuum of states with a single bound photon (shaded in green) and the two-photon continuum (shaded in purple). In this region, which extends up to a coupling strength of about $g/(2J)\simeq 3$ scattering processes of the form 
\begin{equation}
|2_{\rm in}\rangle \leftrightarrow | 1_{\rm out}\rangle |1_{\rm bound}\rangle, 
\end{equation}
are energetically allowed, meaning in particular that scattering processes where two incoming photons evolve into a bound photon and an outgoing one are allowed. Such processes have previously been observed in numerical studies \cite{Longo,Longo2,mistery} and further investigated in Refs.~\cite{Schneider,Koc}. The energy level diagram shown in Fig.~\ref{2photspfig} provides simple energetic arguments to determine under which conditions such processes can occur. Note that all the qualitative considerations so far can be extended to the $N_e$-excitation subspace. For example the $N_e=3$ band structure consists of three-photon continuum of width $12J$, two bands of one bound and two free photons of width $8J$, two bands with two bound and one free photons of width $4J$ and two true three-photon bound states, and so on. Therefore, the complete energy spectrum of a single atom waveguide QED system can be constructed from the knowledge of the $N_e$-photon bound state energies $E_{\pm}^{(N_e)}$.

\subsection{Variational wavefunction}
While the exact eigenstates of the $N_e=2$ subspace can be still found numerically, we now consider a variational approach through which additional intuition about the nature of two-photon bound states can be obtained. In particular, within the two excitation subspace, the lower energy two-photon bound state corresponds to the ground state and can be generically written as
\begin{equation}
\begin{split}
|\Psi^{(2)}_-\rangle=&\left( \cos(\theta)\sigma_+ A_1^\dag -  \sin(\theta)  B_2^\dag\right) |g,0\rangle,
\end{split}
\end{equation}
where $A_1$ and $B_2$ are single- and two-photon operators, respectively.  Based on the discussion in Sec.~\ref{2secp} a suitable ansatz for the two-photon state is 
\begin{equation}
B_2^\dag =  \frac{1}{\mathcal{N}_u} \tilde a^\dag_{\lambda_1} \tilde a^\dag_{\lambda_2},
\end{equation}
where $\tilde a_\lambda= \sum_{x} e^{-\frac{|x|}{\lambda}} a^\dag_x$ and  the normalization constant $\mathcal{N}_u$ is chosen such that $\langle 0|B_2B_2^\dag|0\rangle=1$.  This two-photon wavepacket is an exact solution of the Schr\"odinger equation for $x,y\neq 0$ with an energy
\begin{equation}\label{cos_rel}
E_{-}^{(2)}= -2J \cosh(1/\lambda_1)-2J \cosh(1/\lambda_2).
\end{equation}
For the single-photon operator we demand that the wavefunction also satisfies the first boundary condition, Eq.~\eqref{eqond}, at $x=0$ and $y \ne 0$. This leads to
\begin{equation}
A_1^\dag =  \frac{1}{\mathcal{N}_b} \left[ \sinh\left(\frac{1}{\lambda_2}\right)  \tilde a^\dag_{\lambda_1}  + \sinh\left(\frac{1}{\lambda_1}\right)   \tilde a^\dag_{\lambda_2} \right],
\end{equation}
where $\mathcal{N}_b$ is again a normalization constant. By using this ansatz we can now find an upper bound for the two-photon bound state by minimizing $E_{\rm var}=\langle \Psi^{(2)}_-| H|\Psi^{(2)}_-\rangle$ with respect to $\theta$ and $\lambda_{1,2}$. To further reduce the parameter space, it is reasonable to assume that the wavepacket size of the first photon, $\lambda_1$ is approximately given by the value of $\lambda_-$, which we determined for the single-photon bound state in Sec.~\ref{sec1at}. The variational ansatz is then based on the physical picture of a two-photon dressed state consisting of the single-photon dressed state plus an additional photon, which is more weakly bound and thus less localized, $\lambda_2>\lambda_1$.  As we will show in more detail in a moment, this ansatz provides very accurate values for the bound-state energies.

\begin{figure}
 \includegraphics[width=0.48\textwidth]{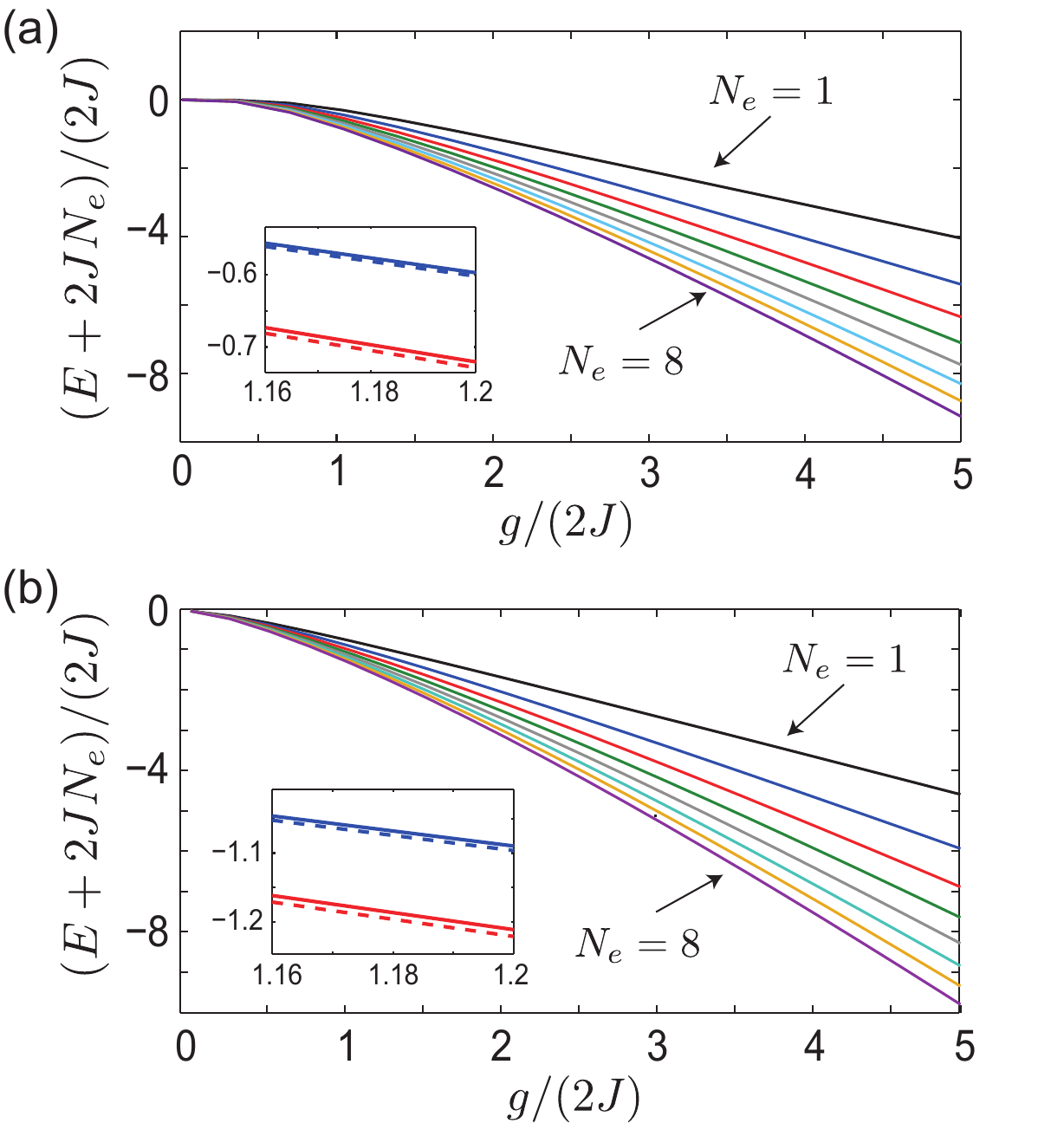}
    \caption{The $N_e$-photon bound-state energies $E_-^{(N_e)}$ obtained from a variational approach are plotted for $N_e=1,\dots,8$ in descending order and for (a) $\delta=0$ and (b) $\delta=-2J$. The dashed lines in the insets show the exact numerical results for $N_e=2$ and $N_e=3$. }
 \label{var_num_en}
\end{figure}
%

\subsection{Multi-photon dressed states}
An important aspect of our variational wavefunction approach is that it can be extended to higher excitation numbers $N_e$ in a systematic way. To do so we write the wavefunction for the lowest energy state within the $N_e$-excitation subspace as
\begin{equation}
\begin{split}
|\Psi_-^{(N_e)}\rangle=&\left( \cos(\theta)\sigma_+ A^\dag_{N_e-1} -  \sin(\theta)  B_{N_e}^\dag\right) |g,0\rangle.
\end{split}
\end{equation}
Based on analogous arguments as above, we make the ansatz
\begin{equation}\label{eq:BNe}
B_{N_e}^\dag =  \frac{1}{\mathcal{N}_u}\tilde a^\dag_{\lambda_1}\tilde a^\dag_{\lambda_2}\dots\tilde a^\dag_{\lambda_{N_e}},
\end{equation}
and
\begin{equation}
\begin{split}
A_{N_e-1}^\dag = & \frac{1}{\mathcal{N}_b}\left[ \sinh{\left(\frac{1}{\lambda_{N_e}}\right)}\tilde a_{\lambda_1}^{\dagger}...\tilde a_{\lambda_{N_e-1}}^{\dagger}...\right.\\
&\left.+\sinh{\left(\frac{1}{\lambda_1}\right)}\tilde a_{\lambda_2}^{\dagger}...\tilde a_{\lambda_{N_e}}^{\dagger}\right],
\end{split}
\end{equation}
where $\mathcal{N}_u$ and $\mathcal{N}_b$ are chosen to normalize each photonic component of the state. To reduce the variational parameter space, the problem can be solved in an iterative manner, i.e., by using the values of $\lambda_1,\dots \lambda_{N_e-1}$ as input for minimizing the energy $E_{-}^{(N_e)}$ with respect to $\theta$ and $\lambda_{N_e}$.

\begin{figure}
 \includegraphics[width=0.48\textwidth]{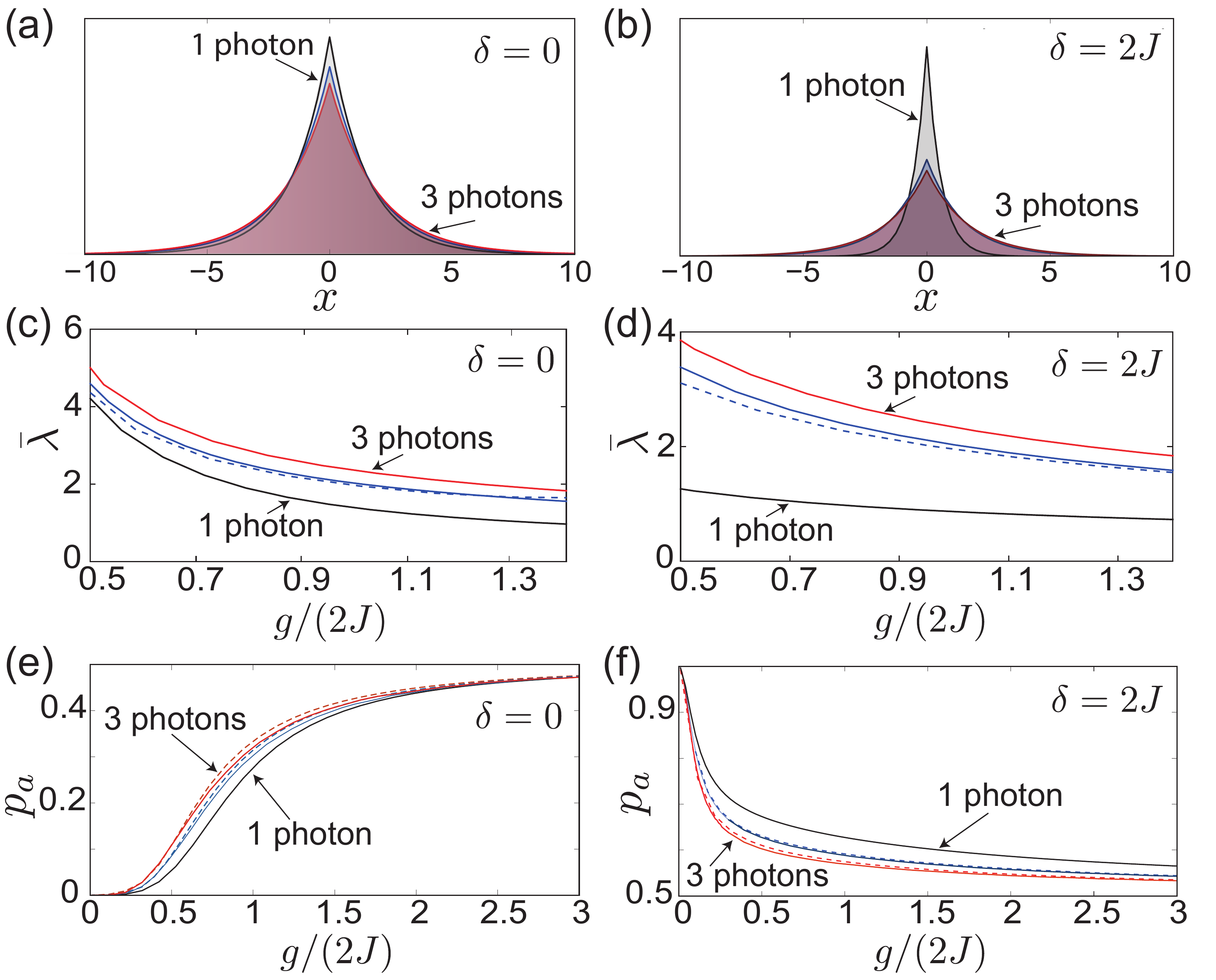}
    \caption{Sketch of the first three photonic wavefunctions that appear in the variational ansatz, Eq.~\eqref{eq:BNe}, for the multi-photon bound states. Here, we have set $g=0.6$, (a) $\delta=0$ and (b) $\delta = -2J-0^+$.  Figures (c) and (d) show the exponential decay length $\bar\lambda_{N_e}$ as a function of $g$, for $N_e=1,2$, and 3 photons and for $\delta=0$ and $\delta = -2J-0^+$, respectively. The dotted line shows the result for $\bar \lambda$ obtained numerically for the case $N_e=2$. In \figs (e) and (f) the atomic population $p_a=\cos^2(\theta(E_-^{N_e}))$ is plotted against $g$ for $\delta=0$ and $\delta = -2J-0^+$, respectively. }
 \label{lambda_var}
\end{figure}

\emph{Discussion.} Figure~\ref{var_num_en} shows the bound-state energies $E^{(N_e)}_{-}$ obtained from our variational approach for up to $N_e=8$ photons. For $N_e=2,3$ these results are compared in the insets with the energies obtained from exact numerical diagonalization in the crossover regime $g/(2J)\sim 1$. The excellent agreement within $\sim1\%$ (for smaller or larger values of $g$ the agreement is even better) demonstrates  that our variational ansatz captures the essential features of the exact wavefunction. 

For $N_e=1,2,3$ the shape of the individual photonic wavepackets associated with the operators $\tilde a^\dag_{\lambda_i}$ in Eq.~\eqref{eq:BNe} are sketched in Fig.~\ref{lambda_var} (a) and (b). We see that in particular near the band edge there is a significant difference between $\lambda_1$ and $\lambda_2$, while the differences between the $\lambda_{N_e}$ are less pronounced for higher excitation numbers.
%
%
It should be noted though that the variational approach, which is constructed to minimize the energy, is not very sensitive to the exponential decay of the wavefunction $\langle 0,\dots,0,x_{N_e}|\Psi^{(N_e)}_-\rangle \sim e^{-|x_{N_e}|/\bar\lambda_{N_e}}$. For physical effects that rely on more accurate predictions for the exponential decay we can, instead of simply setting $\bar\lambda_{N_e}=\lambda_{N_e}$, make use of the exact energy relation
[see Eq.~\eqref{cos_rel} for $N_e=2$]
\begin{equation}\label{eq:coshrelation}
\frac{E_{-}^{(N_e)}}{2J} = \sum_{n=1}^{N_e} \cosh\left(\frac{1}{\bar\lambda_n}\right),
\end{equation}
valid at distances far away from the atom. Therefore, from the exact result for $\lambda_1\equiv \bar\lambda_1$ and the set of bound state energies $E_{-}^{(N_e)}$ obtained from our variational calculations, one can iteratively apply Eq.~\eqref{eq:coshrelation} to also calculate values for the asymptotic decay lengths $\bar\lambda_{N_e}$. For $N_e=2$ the results of this procedure are shown in Fig.~\ref{lambda_var} (c) and (d) and compared with the asymptotic decay length extracted from the numerical solution of the two-photon wavefunction $u(x,y)$. We observe the same general trend as already mentioned above, but at the same time the use of Eq.~\eqref{eq:coshrelation} provides more accurate quantitative results. Finally, in Fig.~\ref{lambda_var} (e) and (f) we plot the atomic population of the $N_e$-photon bound states, showing the expected increase of hybridization for higher excitation numbers.

\begin{figure}
 \includegraphics[width=0.48\textwidth]{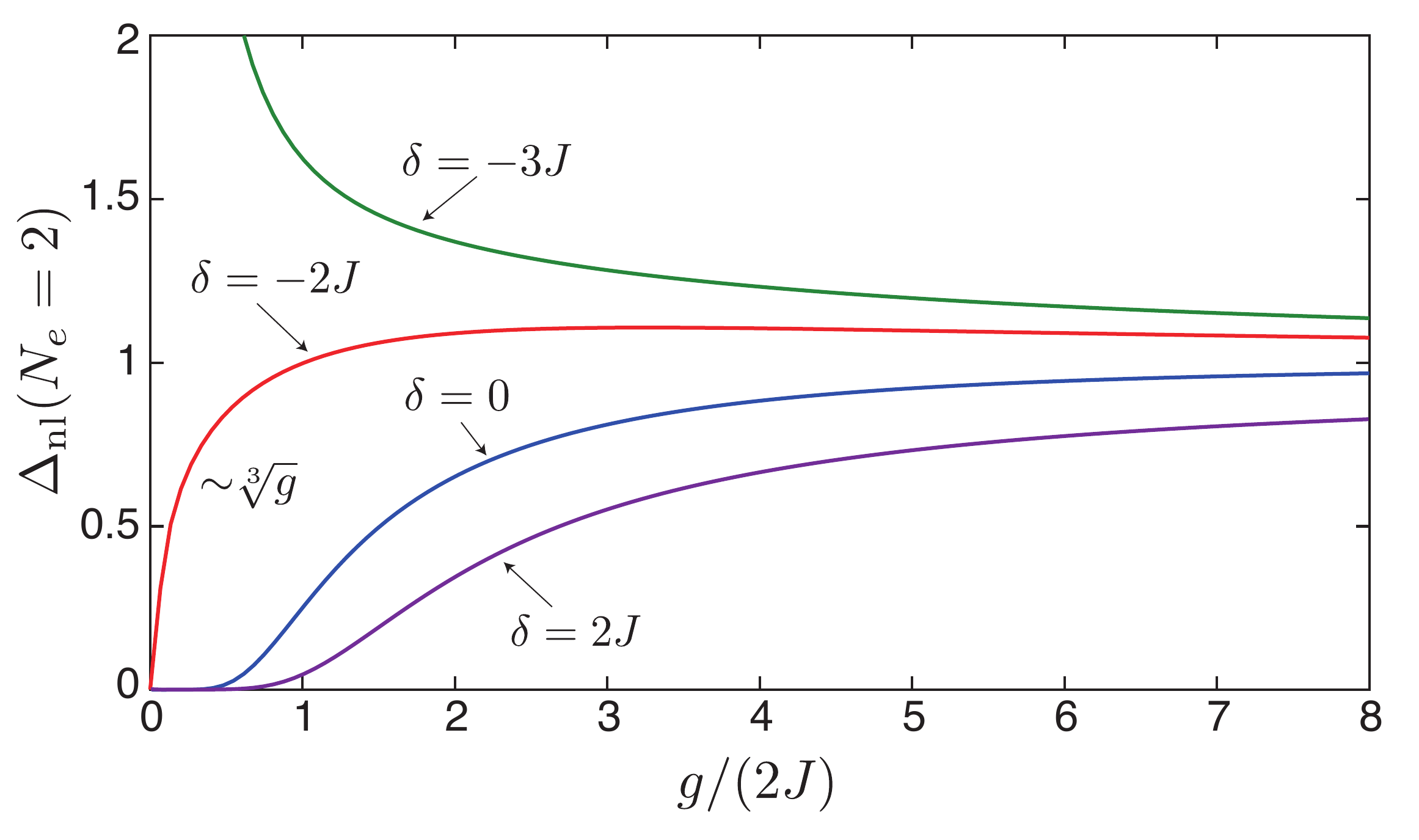}
    \caption{The nonlinearity parameter  $\Delta_{\rm nl}(N_e)$ as defined in Eq.~\eqref{eq:Dnl} is plotted for $N_e=2$ and different atom-photon detunings $\delta$. } \label{plot_nl}
\end{figure}

From Fig.~\ref{var_num_en} we see that for large couplings, $g/(2J)\gg1$, the bound-state energies exhibit  a splitting from the
bare energy by an amount $\sim \sqrt{N_e}$, characteristic of the scaling in conventional cavity QED~\cite{Haroche}. In this limit all bound photons are essentially localized on the atom site and the single-mode physics is recovered. 
To characterize the nonlinearity of the spectrum also in the weak and moderate coupling regime we define the nonlinearity parameter 
\begin{equation}\label{eq:Dnl}
\Delta_{\rm nl}(N_e) =\frac{ |N_eE_{-}^{(1)}-E_{-}^{(N_e)} |}{g|N_e-\sqrt{N_e}|}.
\end{equation} 
With this definition $\Delta_{\rm nl}(N_e)\simeq1$ implies that the excitation spectrum is as nonlinear as cavity QED under resonance conditions, $\delta=0$, while the opposite limit $\Delta_{\rm nl}(N_e)\simeq 0$ indicates a harmonic spectrum.  
In Fig.~\ref{plot_nl}  we plot $ \Delta_{\rm nl}(N_e=2)$ for different values of $g$ and different atomic detunings. We see that, as expected,  in the strong coupling limit, $g\gg \{J, |\delta|\}$, the waveguide QED system approaches asymptotically the nonlinear behavior of the single-mode Jaynes-Cummings model. It can be seen that although for $\delta=-2J$ the nonlinearity (compared to the Jaynes-Cummings nonlinearity) vanishes at small $g$, it is still much stronger than for the resonant case $\delta=0$. This is consistent with the observation that for $\delta=-2J$ the wavelength of the second photon, $\lambda_2$, can be much larger than the wavelength of the first bound photon, $\lambda_1$. In contrast, for $\delta=0$ one finds $\lambda_1\approx \lambda_2$. Note that the approximate scaling of the nonlinearity parameter for $g\rightarrow 0$ can be understood from the simplified assumption $E_-^{(2)} \approx E_-^{(1)} -2J$, which would correspond to  a single photon bound state plus an additional very loosely bound photon at the bandedge. By recalling that $E_-^{(1)}\simeq -2J- [g^4/(4J)]^{1/3}$ (see Eq.\eqref{2JEc}) we obtain $\Delta_{\rm nl}(2)\sim \sqrt[3]{g}$.
For $\delta=-3J$, which for $g\rightarrow 0$ corresponds to an atom-like state inside the bandgap, the nonlinearity parameter diverges. Note that this divergence is a consequence of the chosen normalization for $\Delta_{\rm nl}(N_e)$ and can again be understood from the approximation $E_-^{(2)}\simeq \delta-2J$ for small $g$. 

\section{Dipole-dipole interactions between dressed states}\label{multi-atom}

Our analysis so far has focused on the bound states forming around a single atom.  However, a key element of waveguide QED are the photon-mediated  interactions between two or multiple separated emitters. In the weak-coupling regime discussed in Sec.~\ref{sec_weak}, we have identified effective dipole-dipole interactions between individual atoms, which can be {\it long-range} and scale like $U_{ij}\sim g^2/J$. In the following section we are interested in the corresponding  interactions between dressed states, which represent the elementary waveguide excitations in the strong coupling regime. For previous work on bare atom-atom interactions near band-structures see, for example, Refs.~\cite{Bay,John4,Kuri,GarciaRipollPRA2015}.

\subsection{Two-atom dressed states}

We first consider the case of two atoms located at positions $x_1$ and $x_2$ and focus on the single excitation subspace, $N_e=1$.  
In this case the Schr\"odinger equation can still be solved exactly and details are summarized in App.~\ref{appC}. The resulting energy spectrum has up to four solutions with energies $E_{\pm,s=e,o}$ outside the waveguide continuum given by the real solutions of 
\begin{equation}\label{eig2e}
E_{\pm,e}-\delta=\frac{g^2e^{-\frac{|x_1-x_2|}{2\lambda}}\!\cosh{\lk(\tfrac{|x_1-x_2|}{2\lambda}\rk)}}{E_{\pm,e} \sqrt{1-\frac{4J^2}{E_{\pm,e}^2}}}
\end{equation}
for the even parity states and 
\begin{equation}\label{eig2e}
E_{\pm,o}-\delta=\frac{g^2e^{-\frac{|x_1-x_2|}{2\lambda}}\!\sinh{\lk(\tfrac{|x_1-x_2|}{2\lambda}\rk)}}{E_{\pm,o} \sqrt{1-\frac{4J^2}{E_{\pm,o}^2}}}
\end{equation}
for the odd parity states, where $\lambda\equiv \lambda(E_{\pm,s})$ has the same energy dependence as for the single atom case in Eq.~\eqref{lambda}.

For concreteness and notational simplicity, we restrict the following discussion to the two lower bound states with energies $E_{-,s} <-2J$ below the continuum and even $(s=e)$ or odd $(s=o)$ symmetry of the atom-field system. The corresponding eigenstates can be written as
 \begin{equation}\label{2state}
 |\phi_{s=e,o}\rangle = \frac{1}{\sqrt{2}} \left[ D_{s}^\dag(x_1)  \pm D_{s}^\dag(x_2)\right] |g_1,g_2,0\rangle,
 \end{equation}
where the $+$ (-) sign holds for the state with even (odd) symmetry.  The dressed-state creation operators $D_{e,o}^\dag(x_i)$ are defined as
 \begin{equation}\label{2atoper}
 D_{s=e,o}^\dag(x_i) = \cos(\theta_s) \sigma_+^i + \sin(\theta_s) \frac{\tilde a^{\dag}_{\lambda,s}(x_i)}{\mathcal{N}_{s}},
 \end{equation}
 where  $\tilde a^{\dag}_{\lambda,s}(x_i)=\sum_x e^{-\frac{|x-x_i|}{\lambda}} a^\dag_x$  is an unnormalized photonic creation operator and 
 \begin{equation}\label{2norm}
 \mathcal{N}_{e,o}=\sqrt{\coth{\frac{1}{ \lambda}}\left(1\pm e^{-\frac{|x_1-x_2|}{\lambda}}\right)\pm |x_1-x_2|e^{-\frac{|x_1-x_2|}{\lambda}}},
 \end{equation}
is the corresponding normalization constant [again, the $+$ ($-$) sign holds for the even (odd) case]. The mixing angle $\theta$ is given by 
\begin{equation}\label{2theta}
\cos{\theta_s}=\left(1+\frac{g^2\mathcal{N}^2_{s}}{4J^2\sinh^2{\frac{1}{\lambda}}}\right)^{-\frac{1}{2}},
\end{equation}
which depends on both the bound-state energy and the distance between the atoms.

\begin{figure}
 \includegraphics[width=0.48\textwidth]{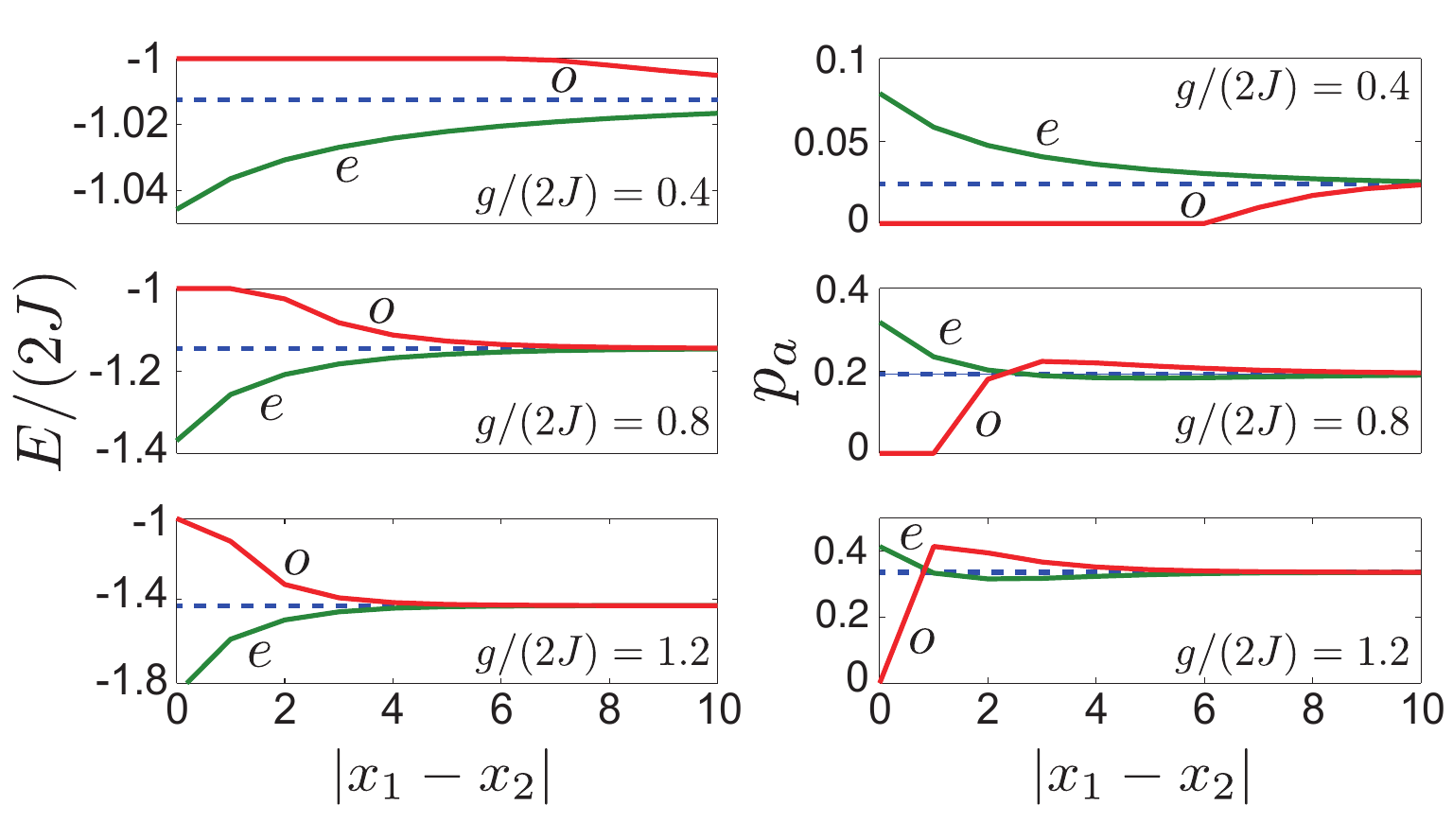}
  \caption{The bound-state energy levels $E_{-,s}$ (left-column panels) and the corresponding atomic populations $p_a=\cos^2(\theta(E_{-,s}))$ (right-column panels) are plotted as a function of the interatomic distance for the case of two atoms and for three representative values of $g/(2J)$. For all plots $\delta=0$ is assumed. For comparison, in each panel the dashed line indicates the corresponding bound-state energy or atomic population for a single atom.}
 \label{fig2aten}
\end{figure}

\emph{Discussion.} Figure \ref{fig2aten} shows the dependence of the two-atom dressed state energies $E_{-,s}$ on the atomic separation $|x_1-x_2|$. For distances which are large compared to $\lambda$ both energies are approximately equal to the single-atom bound state, $E_{-,e}\simeq E_{-,o}\simeq E_{-}$, i.e., there are no long-range interactions. At a large but finite separation $|x_1-x_2|\gtrsim \lambda(E_{-})$ the photonic wavefunctions associated with the single-atom bound states start to overlap so as to induce a splitting of the energies such that $E_{-,e}< E_{-}<E_{-,o}$. As long as this splitting is still small the dressed-states dynamics can be described by the Hamiltonian
\begin{equation}\label{eq:Hdd}
H\approx  \sum_{i=1,2}  E_{-} D^\dag_iD_i  +  \frac{U_{\rm dd}}{2} \left( D^\dag_1 D_2+ D_1 D^\dag_2\right).
\end{equation}  
Here the $D_i\equiv D(x_i)$ are the single-atom dressed state operators introduced in Eq.~\eqref{boundcav}, which 
in the approximated model in Eq.~\eqref{eq:Hdd} are treated as independent, i.e., mutually commuting degrees of freedom.
Therefore, Hamiltonian~\eqref{eq:Hdd} describes a dipole-dipole like coupling between distant dressed states with strength 
(assuming $\delta=0$)
\begin{equation}
U_{\rm dd}\simeq  J \left(  \frac{\cosh{\frac{1}{\lambda}}}{1+\coth^2{\frac{1}{\lambda}}} \right)  e^{- \frac{|x_1-x_2|}{\lambda}}.
\end{equation}
This shows that the long-range interactions occurring in the weak-coupling regime become exponentially localized when $g/(2J)\gtrsim 1$, even in the absence of losses.

 \begin{figure}
 \includegraphics[width=0.48\textwidth]{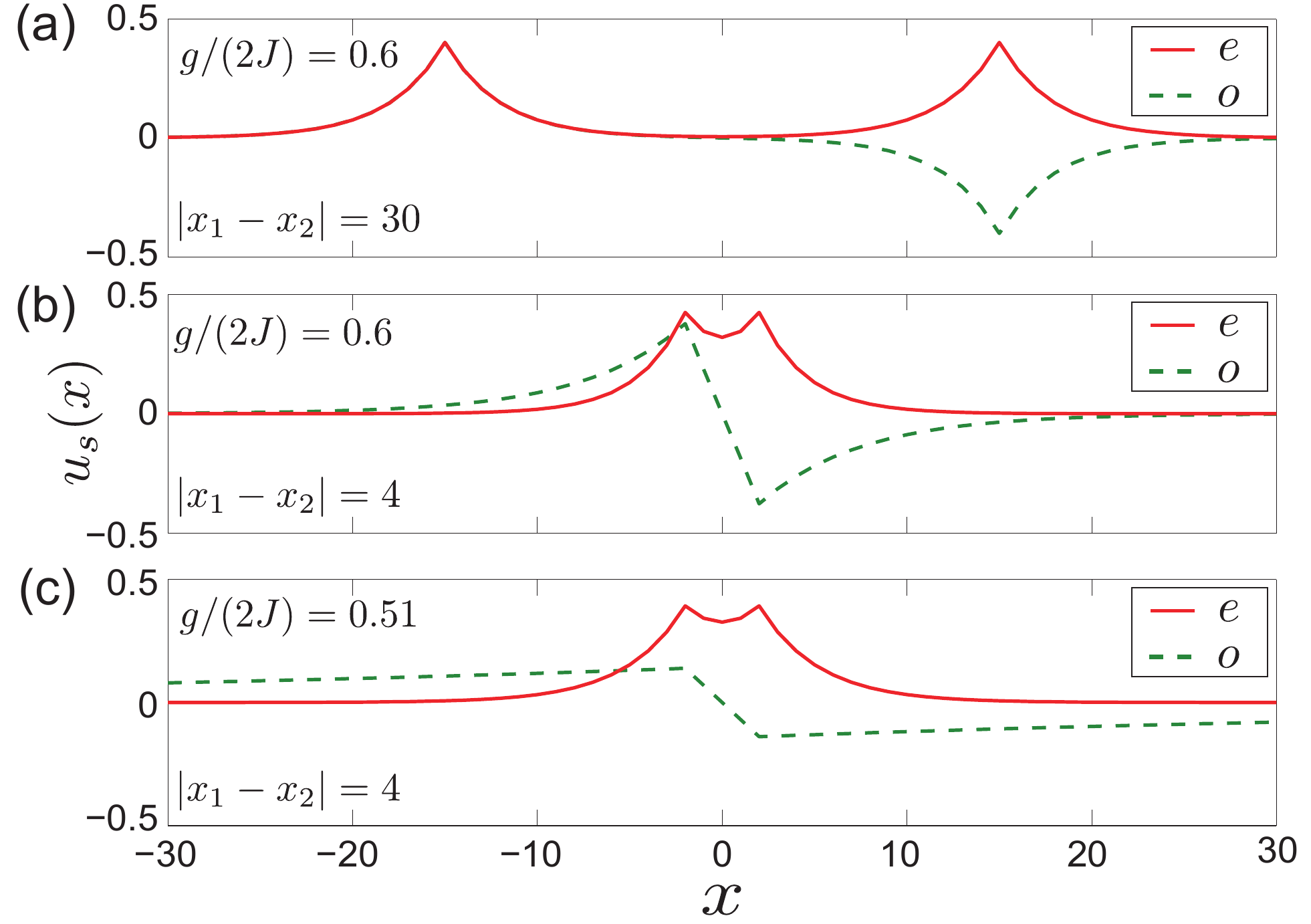}
  \caption{Spatial profile of the photonic wave function $u_s(x)=\langle x|\Phi_s\rangle$ corresponding to the even (red solid line) and odd (green dashed line) lower band  bound states in the case of two atoms for different coupling strengths and interatomic distances. 
For all plots $\delta=0$ is assumed.}
 \label{fig2atwav}
\end{figure}

As the atom-atom separation decreases further, the mutual distortion of the wavepackets must be taken into account. As illustrated in Fig.~\ref{fig2atwav}, the even bound state -- corresponding to the lower level $E_{-,e}$ -- is a `bonding' state such that the photon becomes more and more localized between the atoms. In contrast, the odd state -- corresponding to the upper level $E_{-,o}$ -- behaves as an `anti-bonding' state such that the photon becomes more and more delocalized as the atomic spacing decreases. As a result, two regimes must be distinguished. As shown in more detail in Appendices~\ref{appC} and~\ref{appD}, for $g> g_m$ and $\delta>-2J$, where 
\begin{equation}\label{thresh-up}
g_m= 2J\sqrt{1+\frac{\delta}{2J}}\mbox{,}
\end{equation} 
both $E_{-,e}$ and $E_{-,o}$ solutions exists for all $|x_1-x_2|\geq 1$.  In the opposite case, $g<g_m$, we find that there is a finite distance $x_m=(g_m/g)^2>1$ below which the upper bound state $E_{-,o}$ reaches the band edge and disappears (see Fig. \ref{fig2aten}). This `melting' of  one of the bound states into the waveguide continuum is related to a progressive
delocalization of the photonic wavepacket that eventually
becomes completely delocalized along the array [see for instance the dashed green line in Fig. \ref{fig2atwav}(c)]. This effect is most relevant for resonantly coupled atoms, $\delta\approx 0$, and for moderate coupling strengths, while for $\delta \leq -2J$ both the two-atom bound states always exist. Note that the current discussion has been restricted to the two lower dressed states $E_{-,s}{<}-2J$, but analogous results are obtained for the two-atom bound states above the photonic band, $E_{+,s}>  2J$ with the sign of $\delta$ reversed. See App.~\ref{appD} for more details.

\subsection{Dressed-state bandstructure}\label{bandatomsec}
\begin{figure}
 \includegraphics[width=0.48\textwidth]{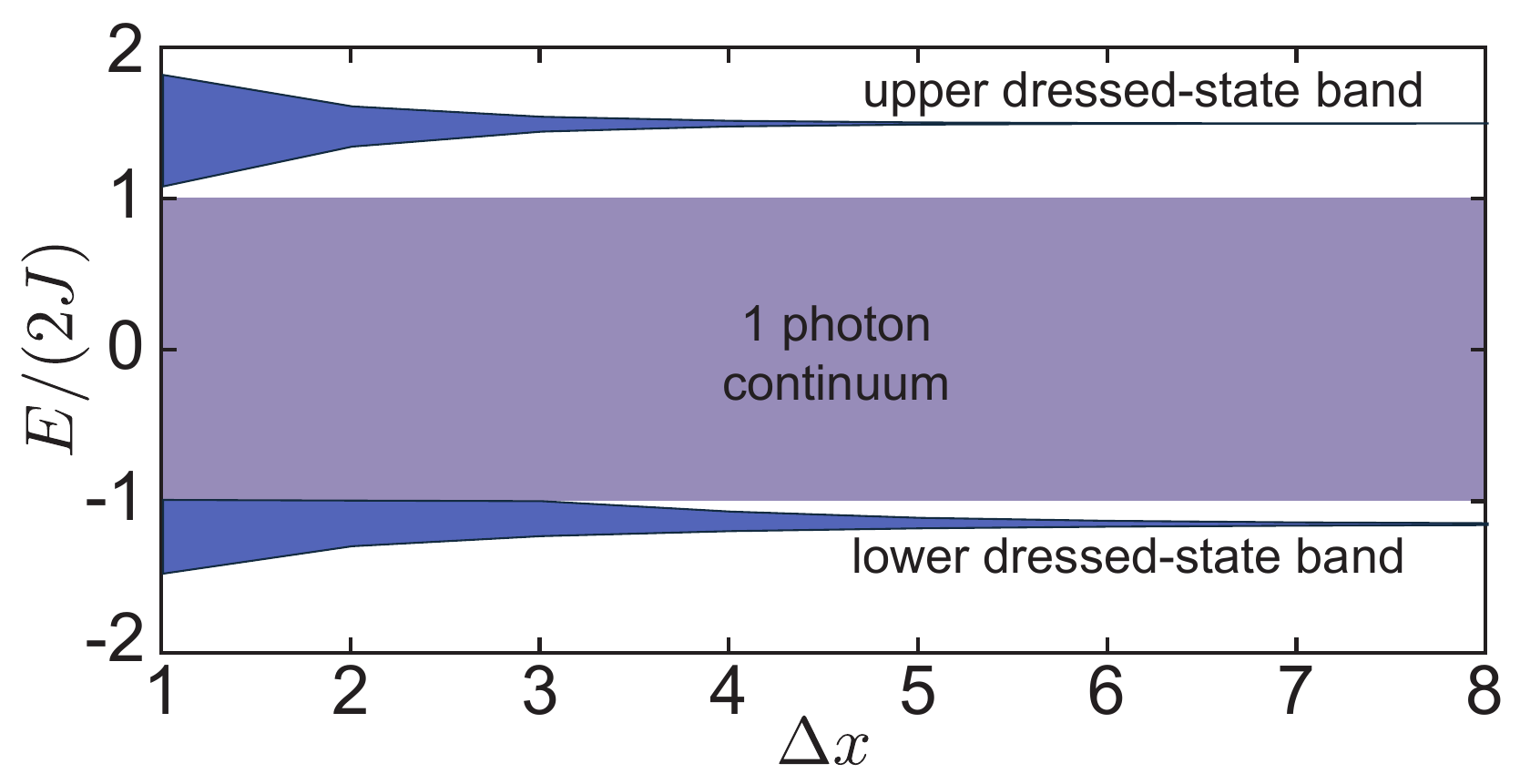}
    \caption{Single-excitation energy spectrum in the case of $N_a=40$ equally spaced atoms as a function of the atomic nearest-neighbour distance $\Delta x$. Note the appearance of upper and lower metabands of bound states. For this plot  $\delta/(2J)=0.6$ and $g/(2J)=1$ have been assumed.}
 \label{multiatomfig}
\end{figure}

The above analysis can be extended to multiple atoms, where for $N\gg N_a\gg1$ and equidistant spacings, $x_{i+1}-x_i=\Delta x$, the coupling between neighboring atoms leads to the formation of a meta-bandstructure for propagating dressed-state excitations below and above the bare photonic band.
This is illustrated in \fig \ref{multiatomfig} where in the single-photon bound-state energies for $N_a=40$ atoms are shown as a function of $\Delta x$. For large $\Delta x$ we see that the bound states form a narrow band around the single-atom energies $E_+$ and $E_-$ with  a width of $\Delta E\approx U_{\rm dd}$. For smaller atomic spacings, the bandwidth grows and -- depending on the parameters -- it can either partially melt into the waveguide continuum or remain energetically separated.
 
As shown in App.~\ref{Nap}, the meta-band is bounded by an upper and lower energy $E_u$ and $E_l$, which obey the equations 
 \begin{equation}\label{Neq1}
E_u-\delta=\frac{g^2\cotha{\lk(\frac{\Delta x}{2\lambda}\rk)}}{E_u \sqrt{1-\frac{4J^2}{E_u^2}}}
\end{equation}
 and 
\begin{equation}\label{Neq2}
E_l-\delta=\frac{g^2\tanha{\lk(\frac{\Delta x}{2\lambda}\rk)}}{E_l \sqrt{1-\frac{4J^2}{E_l^2}}},
\end{equation}
respectively.  Similarly to the previous section, it is possible to define a critical coupling 
\begin{equation}\label{thres many}
 g_m^{(N_a\gg 1)}=\sqrt{2}g_m\,,
 \end{equation}
 for the multi-atom band, which only differs by a factor $\sqrt{2}$ from the two-atom case $g_m$ given in Eq.~\eqref{thresh-up}. For $g>g_m^{(N_a)}$ and $|\delta|/(2J)<1$, the meta-band 
is separated from the phonic continuum regardless of $\Delta x$. In the opposite case, $g<g_m^{(N_a)}$, a fraction of the dressed-state band disappears in the waveguide continuum, i.e., unlike in a usual band-structure only a fraction of the $k$-modes are available.

\section{Conclusions}

In summary, we have analyzed the most essential properties of single-photon, multi-photon and multi-atom dressed-state excitations in a slow-light waveguide QED setup.  Our results provide a both qualitative and quantitative description of the basic linear and nonlinear optical processes in this system and intuitively explain and connect various effects that have been previously described in different limiting cases. We have derived the necessary requirements that are needed to observe atom-photon bound states under realistic experimental conditions, which can be achieved, for example, with state of the art superconducting circuits~\cite{AstafievScience2010,HoiPRL2011,vanLooScience2013,wallraff-superradiance}. More importantly, our analysis of non-linear and multi-atom effects can serve as a starting point to further explored the complexity of waveguide QED systems, when the regime beyond a few excitation is considered.

\acknowledgements
The authors thank Y. Minoguchi, F. Fratini, H. Pichler, F. Lombardo and G. M. Palma for stimulating discussions. This work was supported by the European Project SIQS (600645), the COST Action NQO (MP1403), and the Austrian Science Fund (FWF) through SFB FOQUS F40, DK CoQuS W 1210 and the START grant Y 591-N16. F.C. acknowledges support from Italian PRIN-MIUR  2010/2011MIUR  (PRIN  2010-2011). Work at ICFO has been supported by the ERC Starting Grant FOQAL, the MINECO Plan Nacional grant CANS, and the MINECO Severo Ochoa 
Grant SEV-2015-0522.

\emph{Note added.} After the initial submission of this work a closely related work on multi-photon bound states in waveguide QED systems by T. Shi {\it et al.}~\cite{Tao} appeared.


\appendix

\section{Master equation}\label{app:ME}
In this Appendix we outline the derivation of the master equation \eqref{mastergen} in the weak coupling limit $g/J\rightarrow 0$.  
Starting from Hamiltonian \eqref{totalH} we change into an interaction picture with respect to $H_0= \sum_i \omega_a |e\rangle_i\langle e|+ H_c$ and we obtain the atom-field interaction Hamiltonian 
\begin{equation}
H_{\rm int}(t)= g\sum_{i=1}^{N_a} \!\!\left( \sigma_+^i E(x_i,t) e^{i\omega_a t} + \sigma_-^i E^\dag(x_i,t) e^{-i\omega_a t}Ê \right),
\end{equation}
where
\begin{equation}
E(x,t)= \frac{1}{\sqrt{N}} \sum_k e^{-i \omega_k t} e^{ikx} a_k,
\end{equation}
is the field operator at site $x$ and $k=2\pi m/N$ with $m=-N/2,-N/2+1,...,N/2-1$. The field operators obey the commutation relations 
\begin{equation}\label{comm}
[E(x,t),E^\dag(x',t')]=  \Phi(x-x', t-t'), 
\end{equation}
where 
\begin{equation}\label{bessel}
\begin{split}
\Phi(z, \tau)= &\frac{1}{N}\sum_k e^{-i k z} e^{- i\omega_k \tau} \\
= &\frac{e^{-i\omega_c \tau}}{N}\sum_{n=0}^{N-1} e^{-i 2\pi z n/N} e^{i2J \cos(2\pi n/N) \tau} \\
= &\frac{e^{-i\omega_c \tau}}{N}\sum_{n=0}^{N-1} e^{i 2\pi z n/N} \sum_{m=-\infty}^\infty  i^m J_m(2J\tau)  e^{i 2\pi n m/N}\\
= &e^{-i\omega_c \tau  } i^{|z|} J_{|z|}(2J\tau).
\end{split} 
\end{equation}
Up to second order in $g$ and by performing the usual Born-Markov approximation \cite{petruccione}, we end up with a time-local master equation governing the time evolution of the atom's reduced density operator  
\begin{equation}
\dot \rho(t) = - \int_0^\infty d\tau \, {\rm Tr}_c \{ [H_{\rm int}(t),[H_{\rm int}(t-\tau),\rho_c\otimes \rho(t)]]\},\label{ME}
\end{equation} 
where in the absence of any driving fields $\rho_c=|{0}\rangle\langle {0}|$ is the vacuum state of the waveguide modes. The master equation can be expressed in the form
\begin{equation}
\dot \rho = \sum_{ij} A_{ij} \left(\sigma_-^j\rho \sigma_+^i- \sigma_+^i\sigma_-^j\rho\right) + A^*_{ij} \left(\sigma_-^i\rho \sigma_+^j- \rho \sigma_+^j\sigma_-^i\right), 
\end{equation}
where 
\begin{equation}\label{eq:AijApp}
\begin{split}
A_{ij}= & g^2 \int_0^\infty d\tau \, \langle E(x_i,t) E^\dag(x_j, t-\tau)\rangle e^{i\omega_a \tau}Ê\\
        = & g^2 \int_0^\infty d\tau \, \Phi(x_i-x_j, \tau) e^{i\omega_a \tau}Ê e^{-\gamma_c \tau/2}\\
        = & g^2  i^{|x_i-x_j|} \int_0^\infty d\tau \, J_{|x_i-x_j|}(2J\tau) e^{- \left(\frac{\gamma_c}{2}-i\delta\right) \tau}
\end{split}
\end{equation}
and the cavity decay rate $\gamma_c$ appears through the replacement $\omega_c \rightarrow \omega_c -i\gamma_c /2$.
The final integral can now be evaluated with the help of
\begin{equation}
\int_0^\infty d\tau \, J_{m}(a\tau) e^{-b\tau}Ê= \frac{1}{\sqrt{a^2+b^2}} \left( \frac{a}{b+\sqrt{a^2+b^2}}\right)^m,
\end{equation}
and we obtain
\begin{equation}\label{Aij}
A_{ij}= \frac{g^2 e^{ i K  |x_i-x_j]}}{\sqrt{4J^2-\left(\delta+i\frac{\gamma_c}{2}\right)^2}},  
\end{equation}
where $K$ is given in Eq.~\eqref{eq:K}.
Finally, since $A_{ij}=A_{ji}$ we can regroup the individual terms into the form given in Eq.~\eqref{mastergen}, where we identify  $\Gamma_{ij}= 2 {\rm Re} \{A_{ij}\}$ and $U_{ij}=2{\rm Im}Š\{A_{ij}\}$. 

The derivation of the master equations relies on the validity of the Born-Markov approximation, which requires that the kernel in Eq.~\eqref{eq:AijApp} either decays faster or oscillates faster than the system evolution time set by the coupling $\sim g$. For a single atom this condition  is satisfied as long as $g\ll |\tilde v_g(\delta)|$ and by assuming in addition that $\gamma_a\ll |\tilde v_g(\delta)|$, we can also add to $\Gamma_{ii}$ the bare atomic decay, without influencing the coupling to the waveguide.

For multiple atoms the Bessel function $J_{|x_i-x_j|}(2J\tau)$ reaches its maximum at a finite time
\begin{equation}
\tau \approx \frac{|x_i-x_j|}{2J},
\end{equation}
which reflects the minimal time it takes a photon to propagate between the atoms.
More generally, for the validity of a time-local master equation for $N_a$-atoms with spacing $\Delta x$ we must ensure that the maximal retardation time $\tau_R\sim (N_a-1)\Delta x /|\tilde v_g(\delta)|$ is short compared to the system evolution determined by the single-atom spontaneous-emission time $\Gamma^{-1}$ with $\Gamma=2g^2/|\tilde v_{\rm g}(\delta)|$ [see Eqs. \eqref{Gij} and \eqref{Aij}]. This yields
\begin{equation}
g\ll \frac{|\tilde v_{\rm g}(\delta)|}{\sqrt{(N_a{-}1)\Delta x}}
\end{equation}
as a slightly more stringent condition for large systems. See also Ref.~\cite{moreno,chang-me}.

\section{Single-photon bound states with a single atom}\label{appB}
 In this Appendix we review the derivation of the eigenvalue equation \eqref{eq:Ebound} for the bound states in the case $N_e=1$ and a single atom located at position $x_a$. In particular, this will provide the basis to derive the analogous results in the multi-atom case.
In a frame rotating with frequency $\omega_c$, Hamiltonian \eqref{totalH} can be expressed in the momentum space as
\begin{equation}\label{Htotalk}
\begin{split}
 H {=}&-2J \sum_{k}\cos(k)  a^{\dagger}_ka_k + \sum_{n=1}^{N_a} \delta  |e\rangle_n \langle e|\\
&+\frac{g}{\sqrt{N}}\sum_{n=1}^{N_a}\sum_{k}\left(a_k^{\dagger}\sigma_{-}^ne^{i kx_n}{+}a_k\sigma_+^ne^{-ikx_n}\right).
 \end{split}
 \end{equation}
A state in the single-excitation sector has the form (we set $\sigma_\pm\equiv\sigma_\pm^1$)
\begin{equation}
|\phi\rangle=\left(b\,\sigma_+ +\sum_k c_k a_k^{\dagger}\right)|g,0\rangle\,.
\end{equation}
Plugging this ansatz into the Schr\"{o}dinger equation $H|\phi\rangle=E|\phi\rangle$ yields the coupled equations
\begin{equation}\label{sistbu1}
\begin{split}
&b(E-\delta)=\frac{g}{\sqrt{N}}\sum_k c_k\,e^{-i kx_a}, \\
&c_k(E+2J \cos k)=\frac{g}{\sqrt{N}}\,b\,e^{i kx_a}.
\end{split}
\end{equation}
Using the second equation to eliminate $c_k$ in the first one, we end up with
 \begin{equation}\label{EI0}
E-\delta=\Sigma_1(E)\,,
\end{equation}
where the self-energy $\Sigma_1(E)$ (in the continuous limit) is given by
 \begin{equation}\label{self}
\Sigma_1(E)=\frac{1}{2\pi}\int_{-\pi}^{\pi}\!\!dk\,\frac{g^2}{E+2J\cos k}=\frac{g^2}{E\sqrt{1-\frac{4J^2}{E^2}}}\,,
\end{equation}
where in the last identity we calculated the integral explicitly using that $|E|>2J$ \cite{Eco}. Replacing the self-energy in \eq(\ref{EI0}) we end up with \eq(\ref{eq:Ebound}) in the main text. This equation has two real solutions $E_{\pm}$, where $E_+$ ($E_-$) lies above (below) the continuum $E\in[-2J,2J]$.
The corresponding bound states can be worked out with the help of \eq \eqref{sistbu1} as
\begin{equation}
|\phi_{\pm}\rangle=b(E_{\pm})\left[ \sigma_+ {+}\frac{1}{\sqrt{N}}\sum_k \frac{ge^{i kx_a}}{E_{\pm}+2J\cos{k}}a_k^{\dagger}\right]|g,0\rangle\mbox{,}
\end{equation} 
where, using that the state must be normalized, 
\begin{equation}
b(E)=\left(1+\tfrac{g^2}{E^2\left(1-\frac{4J^2}{E^2}\right)^{\frac{3}{2}}}\right)^{-\frac{1}{2}}\mbox{.}
\end{equation}
In the real space, the bound state reads
\begin{equation}
|\phi_{\pm}\rangle=b(E_{\pm})\left[\sigma_+ +\frac{g \sum_x(\mp 1)^{|x-x_a|}  e^{-\frac{|x-x_a|}{\lambda}}a_x^{\dagger}}{E_{\pm}\sqrt{1{-}\frac{4J^2}{E_{\pm}^2}}}\rk]|g,0\rangle.
\end{equation} 
Exploiting again the normalization of $|\phi_{\pm}\rangle$, one eventually ends up with \eq\eqref{boundcav} defined in terms of the photonic operators $a^\dag_\lambda$ and the mixing angle $\theta$, defined in Eqs.~\eqref{alop} and~\eqref{tetha}, respectively.

\section{Single-photon bound states with many atoms}\label{appC}

For $N_e=1$, but considering multiple atoms the bound states can be derived by exploiting the mirror symmetry of the system.
For the sake of argument, here we focus on bound states below the continuum, i.e., such that $E<-2J$~\cite{note-xn}.
In accordance with the mirror symmetry, we define the pair of collective atomic operators 
\begin{equation}\label{collective}
\mathcal{S}_{s=e,o}=  \sum_{n=1}^{N_a}(\pm 1)^{|n+1|} \sigma_-^n\,,
\end{equation} 
where the + (-) sign holds for $s=e,o$. 
In the case $N_a=2$, the operators \eqref{collective} reduce  to the (unnormalized) symmetric and antisymmetric combinations of $\sigma_-^1$ and $\sigma_-^2$. Based on this definition, here we look for bound states of the form
\begin{equation}\label{phisNa}
|\phi_{s}^{(N_a)}\rangle=\left(b\,\mathcal{S}_{ s}^{\dagger} +\sum_k c_k a_k^{\dagger}\right)|g,{\dots},g,0\rangle\,.
\end{equation}
If $N_a>2$, the bound states defined in Eq.~\eqref{phisNa} are those whose energies form the boarders of the dressed-state metabands (see Fig.~ \ref{multiatomfig}).  
Imposing the ansatz \eqref{phisNa} to be an eigenstate of Hamiltonian (\ref{Htotalk}) with eigenvalue $E$ yields
an eigenvalue equation analogous to Eq.~\eqref{EI0}  with the self-energy now given by 
\begin{eqnarray}\label{self-many}
\Sigma_{s}(E)&=&\sum_{n}(\pm 1)^{|n+1|}\frac{1}{2\pi}\int_{-\pi}^{\pi}\!\!dk\,\frac{g^2 e^{ik (x_n-x_a)}}{E+2J\cos k}\nonumber\\
&=&\Sigma_1(E)\,\,f_{N_a,s}(E)\,,
\end{eqnarray}
where $\Sigma_1(E)$ is the single-atom self-energy in Eq.~\eqref{self} and
\begin{equation}\label{f-function}
f_{N_a,s}(E)=\sum_n  (\pm 1)^{|n+1|}e^{-\frac{ |x_n-x_a|}{\lambda}},
\end{equation}
with $\lambda=\lambda(E)$ being the same energy function as in Eq.~\eqref{lambda}. We introduced the atomic position $x_a$ that set the choice of placing the atomic ensemble in the array. As in the one-atom case, in deriving the last identity of Eq.~\eqref{self-many} we used $E<-2J$ to calculate
the integral over $k$ through standard methods \cite{Eco}.

The self-energy, hence the eigenvalue equation, is thus determined by the function $f_{N_a,s}(E)$ in Eq.~\eqref{f-function}.  We will analyze this function now in more detail for the paradigmatic cases $N_a=2$ and $N_a\gg1$, which are the cases considered in Sec.~\ref{multi-atom}.

\subsection{Two atoms}

For $N_a=2$ and choosing $x_a=x_1$, Eq.~\eqref{f-function} simply yields 
\begin{equation}\label{f2}
f_{2,e}=e^{-\frac{\Delta x}{2\lambda}}\!\cosh{\lk(\tfrac{\Delta x}{2\lambda}\rk)}\,,\,\,\,f_{2,o}=e^{-\frac{\Delta x}{2\lambda}}\!\sinh{\lk(\tfrac{\Delta x}{2\lambda}\rk)},
\end{equation}
for the even- and odd-parity states, respectively (recall that $\Delta x=|x_1{-}x_2|$). 
This provides the self-energy and thus the eigenvalue equation for the energies $E_{-,s}$ [see \eq (\ref{self-many})].
The corresponding bound states can be derived in terms of $E_{-,s}$ in a way essentially analogous to that in Appendix \ref{appB}.
For bound states below the continuum, this gives
\begin{equation}\label{phims}
\begin{split}
&|\phi_{-,s}\rangle=b(E_{-,s})\left[ \sigma_+^1 \pm \sigma_+^2\right.\\
&\left. \,\,\,\,\,\,\,\,\,\,\,\,\,\,\,\,\,\,\,\,+\frac{1}{\sqrt{N}}\sum_k \frac{g( e^{ik x_1}\pm  e^{ik x_2})}{E_{-,s}+2J\cos k}a_k^{\dagger}\right]|g_1,g_2,0\rangle\,,
\end{split}
\end{equation} 
where function $b(E)$ follows from the normalization constraint and reads
\begin{equation}
b(E)=\left(2+\frac{g^2\mathcal{N}^2_{s}}{2J^2\sinh^2{\frac{1}{\lambda}}}\right)^{-\frac{1}{2}},
\end{equation} 
with $\mathcal{N}_{s}$ defined by \eq(\ref{2norm}). In position space, state (\ref{phims}) reads
\begin{equation}
\begin{split}\label{x-space}
&|\phi_{-,s}\rangle=b(E_{-,s})\Big[ \sigma_+^1 \pm \sigma_+^2
{+}\frac{g}{E_{-,s}\sqrt{1-\frac{4J^2}{E_{-,s}^2}}}\\
&\,\,\,\,\,\,\,\,\,\,\,\,\,\,\,\,\,\,\,\,\,{\times} \sum_x\! \left( e^{-\frac{|x-x_1|}{\lambda}}\pm e^{-\frac{|x-x_2|}{\lambda}}\right)a_x^{\dagger}\Big]|g_1,g_2,0\rangle \mbox{.}
\end{split}
\end{equation} 
In analogy with the single-atom case, one can arrange such bound states in the form \eqref{2state} in terms of the polaritonic operators \eqref{2atoper} and the mixing angle \eqref{2theta}.

Regarding bound states above the band, one can follow an analogous reasoning by taking into account the different definition of operators $\mathcal{S}_s$ \cite{note-xn}. While this affects the expression
of the bound states, namely the counterparts of \eqs(\ref{phims}) and (\ref{x-space}), \eq (\ref{self-many}) for the self-energy turns out to be unaffected. The self-energies (\ref{self-many}) thereby hold
both above and below the continuum. 

At this time, we also mention that -- while our approach based on the collective atomic operators (\ref{collective}) is devised so as to easily tackle the $N_a\gg1$ limit -- in the $N_a=2$ case an equivalent method would be to block-diagonalize $H$ with the blocks corresponding to even- and odd-parity sectors of the entire single-excitation Hilbert space (including the field). In the even (odd) subspace, the problem is reduced to an effective single atom coupled to the cosine-shaped (sine-shaped) field modes. This approach was followed in Ref.~\cite{ordonez}, where however the authors focused on bound states {\it in} the continuum (BIC)~\cite{bic} only. The effective Hamiltonian in each parity-definite subspace differs from the Fano-Anderson model in Eq. \eqref{Htotalk} (case $N_a=1$) in that the atom-mode couplings are $k$-dependent. Such ``coloured" Fano-Anderson model was first investigated in Ref.~\cite{Longhi} in the case of sine-shaped couplings.

 \subsection{ $N_a\gg1$ atoms}\label{Nap}
 
In the limiting case of a very large number of equispaced atoms, $N_a\gg1$, function (\ref{f-function}) can be written in a compact form, by setting $x_a=x_{N_a/2}$~\cite{note-Na}, since it reduces to a geometric series.
By expressing in \eq(\ref{f-function}) each atomic position as $x_n=x_a+(n-N_a/2)\Delta x$, we end up with
\begin{equation}\label{f-many}
f_{N_a \gg1,e}=\cotha{\lk(\frac{\Delta x}{2\lambda}\rk)}\,,\,\,\,f_{N_a\gg 1,o}=\tanha{\lk(\frac{\Delta x}{2\lambda}\rk)}\,,
\end{equation}
which provides the eigenvalue equation for the bound states $|\phi_{s}^{(N_a\gg1)}\rangle$. We confirmed  numerically  that the metaband-edge levels (see Fig.~\ref{multiatomfig}) for growing $N_a$ converge 
to the numerical solutions of the eigenvalue equation $E-\delta=\Sigma_1(E)f_{N_a \gg1,s}(E)$. Specifically, above the continuum ($E>2J$) the solution for $s=e$ ($s=o$) gives the upper (lower) metaband edge,
while below the continuum $s=e$ ($s=o$) corresponds to the lower (upper) metaband edge.

\section{Multi-atom bound-states}\label{appD}

Here we address a number of properties of the multi-atom bound-state levels in the case $N_a=2$ and $N_a\gg1$ with the goal of proving 
the salient features of the energy spectra in Fig. \ref{fig2aten} and \ref{multiatomfig} discussed in the main text.

\subsection{$N_a=2$}

As discussed in App.~\ref{appC}, the bound-state levels are the solutions of the equation $E-\delta=\Delta_s(E)$ in the domain $|E|>2J$.
Using \eqs (\ref{self-many}), (\ref{f2}) and (\ref{lambda}), the self-energy function explicitly reads
 \begin{equation} 
\Sigma_{s}(E){=}\frac{g^2}{E\sqrt{1{-}\frac{4J^2}{E^2}}}\!\!\left[1{\pm}\left(\frac{|E|}{2J}{-}\frac{|E|}{2J}\sqrt{1{-}\frac{4J^2}{E^2}}\right)^{\!\!|x_1{-}x_2|}\right]\!\!,\label{deltapm}
 \end{equation}
 where as usual the $+$ ($-$) sign holds for $s=e$ ($s=o$). The corresponding expression for $E<-2J$ follows straightforwardly from the fact that $\Sigma_s(E)$ is an odd function of $E$. 

{\it Below the continuum}, i.e., for $E<-2J$, both $\Sigma_e(E)$ and $\Sigma_o(E)$ monotonically decrease with $E$ [\cf\eq(\ref{deltapm})]. Thereby, if the value taken by the linear function $y=E-\delta$ at $E=-2J$ lies above $\Sigma_s(-2J)$ then a single bound state (for fixed $s$) of energy $E_{-,s}<-2J$ certainly occurs. This condition thus explicitly reads $-2J-\delta>\Sigma_s(-2J)$.
This is always fulfilled for $s=e$ given that $\Sigma_e(-2J)=-\infty$. Instead, for $s{=}o$, by calculating $\Sigma_o(-2J)=-g^2|x_1-x_2|/(2J)$ [see \eq(\ref{deltapm}) for $E{\rightarrow}(-2J)^+$], the above condition results in
  \begin{equation} 
g>\frac{2J\sqrt{1+\frac{\delta}{2J}}}{\sqrt{|x_1-x_2|}}=\frac{g_m}{\sqrt{|x_1-x_2|}},\label{thresh-above}
\end{equation}
where $g_m$ is the same as in \eq(\ref{thresh-up}).
Hence, as discussed in Sec. \ref{multi-atom}, both $E_{-,e}$ and $E_{-,o}$ solutions exist for any interatomic distance when $g>g_m$. If instead $g<g_m$, at the critical distance $|x_1-x_2|=(g_m/g)^2$ the solution $E_{-,o}$ merges with the continuum, i.e., $E_{-,o}=-2J$, and it no longer exists for $|x_1-x_2|<x_m$ (see \fig \ref{fig2aten}). Moreover, note that in the light of the geometrical criterion given above if $E_{-,o}$ exists then $E_{-,o}>E_{-,e}$ since $\Sigma_o(E)>\Sigma_e(E)$ [\cf\eq(\ref{deltapm})]. 
\eq(\ref{thresh-above}) holds for $\delta>-2J$. For $\delta\leq -2J$, $E_{-,o}$ always exists since $-2J-\delta$ is positive while $\Sigma_o(2J)$ is negative anyway.

As for bound states {\it above the continuum}, a similar reasoning can be carried out. Recalling that $\Sigma_s(-E)=-\Sigma_s(E)$, we have $\Sigma_o(2J)=g^2|x_1-x_2|/(2J)$ and $\Sigma_e(2J)=+\infty$ with both functions $\Sigma_s(E)$ mononically decreasing with $E$ for $E>2J$. The condition for the existence of a bound state will now read $2J-\delta<\Sigma_s(2J)$. Again, it is always fulfilled when $s=e$ since $\Sigma_e(2J)$ diverges to $+\infty$. Instead, for $s=o$ the threshold condition for $\delta<2J$ reads
\begin{equation} 
g>\frac{2J\sqrt{1-\frac{\delta}{2J}}}{\sqrt{|x_1-x_2|}}\,,\label{thresh-below}
\end{equation}
which is analogous to \eq(\ref{thresh-above}) but the replacement $\delta\rightarrow-\delta$ in the expression of $g_m$.
For $\delta >2J$ both levels $E_{+,s}$ exist. Moreover, since now $\Sigma_o(E)<\Sigma_e(E)$ we have $E_{+,o}<E_{+,e}$.

To summarize, outside the continuum, a pair of bound states of even symmetry and energies $E_{\pm,e}$ always exist, one above and one below the photonic band.  At most two further odd-symmetry bound states of energies $E_{\pm,o}$ may be present as well, depending on the values of $g$, $|x_1-x_2|$ and $\delta$. Note that, for $|\delta|<2J$, the critical coupling strengths appearing in \eqs (\ref{thresh-above}) and (\ref{thresh-below}) are different, which entails that three cases are possible: $E_{+,o}$ exists while $E_{-,o}$ does not (or vicecersa), $E_{\pm,o}$ both exist, $E_{\pm,o}$ both do not exist. Combining together \eqs (\ref{thresh-above}) and (\ref{thresh-below}), the conditions for these three cases to occur, for $|\delta|\leq2J$, read
\begin{eqnarray}
&&g>\tfrac{2J\sqrt{1{+}\tfrac{|\delta|}{2J}}}{\sqrt{|x_1-x_2|}}\,\,\Leftrightarrow\,\,\,{\rm both}\,\,\,E_{+,o} \,\,{\rm and} \,\,E_{-,o}\,\, {\rm exist}\,,\label{table3}\\
&&\tfrac{2J\sqrt{1{-}\tfrac{|\delta|}{2J}}}{\sqrt{|x_1-x_2|}}{<}g{<}\tfrac{2J\sqrt{1{+}\tfrac{|\delta|}{2J}}}{\sqrt{|x_1-x_2|}}\,\,\Leftrightarrow\,\,\,{\rm only}\,\,E_{\rm sgn\,(\delta),o} \,\, {\rm exists}\label{table4},\,\,\,\,\,\,\,\,\,\\
&&g<\tfrac{2J\sqrt{1{-}\tfrac{|\delta|}{2J}}}{\sqrt{|x_1-x_2|}}\,\,\Leftrightarrow\,\,\,{\rm neither}\,\,\,E_{+,o} \,\,{\rm nor} \,\,E_{-,o}\,\, {\rm exist}\label{table5}\,.
\end{eqnarray}

\subsection{$N_a\gg1$}

The analysis for $N_a\gg1$ proceeds similarly to the $N_a=2$ case. The explicit self-energy functions $\Sigma_{s=e,o}(E)$ are obtained from \eqs (\ref{self-many}), (\ref{f-many}) and (\ref{lambda}). Like in the 2-atom case,
$\Sigma_{e}(E)>\Sigma_{o}(E)$ [$\Sigma_{e}(E)<\Sigma_{o}(E)$] for $E>2J$ ($E<-2J$) with $\Sigma_{e}(E)$ diverging to $+\infty$ and $-\infty$ for $E\rightarrow (2J)^+$ and $E\rightarrow (-2J)^-$, respectively. Instead, $\Sigma_{o}(\pm 2J)\ug \pm g^2\Delta x/(4J)$. Accordingly, the same geometrical criterion as in the previous subsection entails that the conditions for the existence of $E_{+,o}$ and $E_{-,o}$ are the same as in \eqs (\ref{thresh-above}) and (\ref{thresh-below}), respectively, apart from the factor $\sqrt{2}$ on either right-hand side. The same factor thereby appears in \eqs (\ref{table3})-(\ref{table5}), which are now interpreted as the
conditions for establishing whether none [\eq (\ref{table3})], only one [\eq (\ref{table4})] or both [\eq (\ref{table5})] of the metabands merge with the photonic band.



\begin{thebibliography}{99}
\bibitem{Footnote1} For $N_a>1$, there may also exist localized bound states of energy $|E|< 2J$, so-called bound states in the continuum (BIC) (see Ref.~\cite{bic,Longhi, ordonez}) with a wavefunction that is localized between the atoms due to destructive interference.
\bibitem{note-xn} Above the band, the reasoning is analogous but an extra phase factor $(-1)^{|x_n{-}x_a|}$ needs to be included in the definition of the collective operators (\ref{collective}), i.e., $\Sigma_{s}{=} \sum_{n}(\pm 1)^{|n+1|} (-1)^{|x_n-x_a|}\sigma_-^n$.
\bibitem{note-Na}  An analogous reasoning can be made if $N_a$ is odd.
\end{thebibliography}
\end{document}